\documentclass{lmcs} 
\pdfoutput=1

\usepackage{lastpage}
\lmcsdoi{15}{2}{1}
\lmcsheading{}{\pageref{LastPage}}{}{}%
{Jul.~04,~2017}{Apr.~04,~2019}{}

\usepackage{amsmath}
\usepackage{amssymb}
\usepackage{amsthm}
\usepackage{stmaryrd}
\usepackage{xspace}
\usepackage{algorithm}
\usepackage[noend]{algpseudocode}
\usepackage{mathtools}
\usepackage{hyperref}

\usepackage{graphicx}
\usepackage{xcolor,tikz,pgffor}
\usetikzlibrary{decorations,arrows,automata,positioning}
\tikzset{every state/.style={minimum size=2em,initial text={}}}

\theoremstyle{plain}

\newcommand{\vpt}{\ensuremath{\mathsf{VPT}}\xspace}
\newcommand{\VPT}{\ensuremath{\mathsf{VPT}}\xspace}
\newcommand{\fvpt}{\ensuremath{\mathsf{fVPT}}\xspace}
\newcommand{\fvpts}{\ensuremath{\mathsf{fVPTs}}\xspace}
\newcommand{\vpts}{\ensuremath{\mathsf{VPTs}}\xspace}
\newcommand{\VPTs}{\ensuremath{\mathsf{VPTs}}\xspace}

\newcommand{\wn}{\ensuremath{\mathsf{wn}}\xspace}

\newcommand{\fsts}{\ensuremath{\mathsf{FSTs}}\xspace}
\newcommand{\fst}{\ensuremath{\mathsf{FST}}\xspace}

\newcommand{\h}{\ensuremath{\mathsf{h}}\xspace}
\newcommand{\hc}{\ensuremath{\mathsf{hc}}\xspace}
\newcommand{\ar}{\ensuremath{\mathsf{ar}}\xspace}
\newcommand{\eval}{{\textsc{LcpIn}}\xspace}
\newcommand{\evalTT}{\ensuremath{M_\textsc{\eval}}\xspace}
\newcommand{\factorize}{{\textit{factorize}}\xspace}
\newcommand{\nextChar}{{\textit{next}}\xspace}

\newcommand{\MTP}{MTP\xspace}
\newcommand{\dconf}{d-configuration\xspace}
\newcommand{\dconfs}{d-configurations\xspace}
\newcommand{\lcp}[1]{\ensuremath{\mathsf{lcp}(#1)}\xspace}
\newcommand{\lcpin}[2]{\ensuremath{\mathsf{lcp_{in}}(#1,#2)}\xspace}
\newcommand{\maxoutdiff}{\ensuremath{\mathsf{out}_{\not=}}\xspace}
\newcommand{\reach}{\ensuremath{\mathsf{reach}}\xspace}
\newcommand{\hmid}{\ \mid\ }
\newcommand{\update}{\ensuremath{\mathsf{update}}\xspace}
\newcommand{\EOS}{\dashv}

\newcommand{\Figure}{Fig.}
\newcommand{\str}[2]{G_{#1}^{#2}}
\newcommand{\stredges}[1]{#1.\textit{edges}()} 
\newcommand{\stredgesymb}[1]{\xhookrightarrow{#1}}
\newcommand{\stredge}[3]{{#1}\stredgesymb{#2}{#3}}
\newcommand{\strleaves}[1]{#1.\textit{leaves}()} 
\newcommand{\rootstr}{\#}

\newcommand{\emptyword}{\epsilon}

\newcommand{\Nat}{\mathbb{N}}

\newcommand{\nl}{{\sc NL}\xspace}

\newcommand{\exptime}{{\sc ExpTime}\xspace}

\newcommand{\ptime}{{\sc PTime}\xspace}

\newcommand{\conptime}{{\sc coNPTime}\xspace}

\newcommand{\OBM}{{\sc OBM}\xspace}
\newcommand{\HBM}{{\sc HBM}\xspace}

\newcommand{\del}{\Delta}

\newcommand{\olivier}[1]{}
\newcommand{\ignore}[1]{}

\newcommand{\COMMENT}[2][.5\linewidth]{%
    \leavevmode\hfill\makebox[#1][l]{//~#2}}

\newcommand{\Left}[1]{\ensuremath{\mathsf{left}(#1)}}
\newcommand{\Right}[1]{\ensuremath{\mathsf{right}(#1)}}

\newcommand{\dom}{\textit{Dom}} 

\newcommand\inter[1]{\llbracket#1\rrbracket}

\newcommand{\ie}{i.e.\xspace}

\newcommand{\manu}[1]{}

\newcommand{\newcontent}[1]{\textcolor{blue}{#1}}

\newcommand\newEdges{\textit{newEdges}}
\newcommand\orphans{\textit{orphans}}
\newcommand\removeEdges{\textit{remove\_edges}}
\newcommand\removeLeaves{\textit{remove\_leaves}}
\newcommand\done{\textit{done}\xspace}
\newcommand\add{\textit{add}\xspace}
\newcommand\remove{\textit{removeEdge}\xspace}
\newcommand\pop{\textit{pop}\xspace}

\renewcommand{\newcontent}[1]{#1}

\begin{document}


 \title{Streamability of Nested Word Transductions}

\author[E. Filiot]{Emmanuel Filiot\rsuper{a}}
\address{\lsuper{a}Universit\'e Libre de Bruxelles}

\author[O. Gauwin]{Olivier Gauwin\rsuper{b}}
\address{\lsuper{b}LaBRI, CNRS, Universit\'e de Bordeaux}

\author[P.-A. Reynier]{\texorpdfstring{\\}{} Pierre-Alain Reynier\rsuper{c}}
\address{\lsuper{c}Aix-Marseille Univ, Universit\'e de Toulon, CNRS, LIS, Marseille, France}

\author[F. Servais]{Fr\'ed\'eric Servais\rsuper{d}}
\address{\lsuper{d}Ecole Sup\'erieure d'Informatique de Bruxelles}

\thanks{Partially supported by the ESF project GASICS, by the FNRS,
  by the PAI program Moves funded by the Federal Belgian Government,
  the ANR projects ExStream (ANR-13-JS02--0010--01) and DeLTA (ANR--16--CE40--0007),
  and the FET project FOX (FP7--ICT--233599).}

\keywords{visibly pushdown transducers, streaming, nested words, online algorithms}
\subjclass{F.1.1, F.1.2, F.2.2} 
\titlecomment{A preliminary version of this work has been presented
  at the FSTTCS'11 conference~\cite{FiliotGauwinReynier11}.
  Links with this version are discussed in the introduction.}

\begin{abstract}
  \noindent 
  We consider the problem of evaluating in streaming (i.e., in
  a single left-to-right pass) a nested word transduction with a
  limited amount of memory.  A transduction T is said to be height
  bounded memory (HBM) if it can be evaluated with a memory that
  depends only on the size of T and on the height of the input
  word. We show that it is decidable in coNPTime for a nested word
  transduction defined by a visibly pushdown transducer (VPT), if it
  is HBM\@. In this case, the required amount of memory may depend
  exponentially on the height of the word. We exhibit a sufficient,
  decidable condition for a VPT to be evaluated with a memory that
  depends quadratically on the height of the word. This condition
  defines a class of transductions that strictly contains all
  determinizable VPTs.
\end{abstract}

\maketitle

\section*{Introduction}

Memory analysis is an important tool for ensuring system robustness.
In this paper we focus on the analysis of programs processing
\emph{nested words}~\cite{AlurJACM09}, \emph{i.e.}, words with a recursive structure,
like program traces, XML documents, or more generally unranked trees.
On huge inputs, a \emph{streaming} mode is often used,
where the nested word is read only once, from left to right.
This corresponds to a depth-first left-to-right
traversal when the nested word is considered as a tree.
For such programs,
\emph{dynamic analysis} problems have been addressed in various contexts.
For instance,
runtime verification detects dynamically,
and as early as possible, whether
a property is satisfied by a program trace~\cite{KupfermanVardi01,BauerLeuckerSchallhart11}.
On XML streams, some algorithms outputting nodes
selected by an XPath expression
at the earliest possible event
have also been proposed~\cite{BenediktJeffrey07,GauwinNiehrenTison09b}.
These algorithms allow minimal buffering~\cite{BaryossefFontouraJosifovski05}.

In this paper, we investigate \emph{static analysis} of
memory usage for a special kind of programs on nested words,
namely programs defined by \emph{transducers}. We assume that the
transducers are \emph{functional} and \emph{non-deterministic}.
Non-determinism is required as input
words are read from left to right in a single pass and
some actions may depend on the future of the stream. For instance, the
XML transformation language XSLT~\cite{XSLT} uses XPath
for selecting nodes where local transformations are applied,
and XPath queries relies
on non-deterministic moves along tree axes, such as a move
to any descendant. We require our transducers to be
\emph{functional}, as we are mainly interested by transformation
languages like XSLT~\cite{XSLT},
XQuery~\cite{XQuery}
and XQuery Update Facility,~\cite{XQUF},
for which
any transformation maps each XML input document to a unique
output document.

\emph{Visibly pushdown transducers} (\vpts) form a subclass of pushdown transducers
adequate for dealing with nested words and streaming evaluation, as
the input nested word is processed from left to right. They are
visibly pushdown automata~\cite{AlurJACM09} extended with arbitrary
output words on transitions.
\vpts capture interesting fragments of the aforementioned XML
transformation languages that are amenable to efficient streaming
evaluation, such as all editing operations
(insertion, deletion, and relabeling of nodes, as used for instance
in XQuery Update Facility~\cite{XQUF})
 under all regular tests.
Like for visibly pushdown automata, the stack
behavior of \vpts is imposed by the type of
symbols read by the transducer. Those restrictions on stack operations
allow to decide functionality and equivalence of functional \vpts
in \ptime and \exptime respectively~\cite{vpts10}.

Some transductions defined by (functional and non-deterministic) \vpts
cannot be evaluated efficiently in streaming. For instance, swapping
the first and last letter of a word can be defined by a \vpt as
follows: guess the last letter and transform the first letter into the
guessed last letter, keep the value of the first letter in the state,
and transform any value in the middle into itself. Any deterministic
machine implementing this transformation
requires to keep the entire word in memory until the last letter is read.
It is not reasonable in practice as for instance XML documents can be very
huge.

Our aim is thus to identify decidable classes of transductions
for various memory requirements that are suitable to space-efficient streaming
evaluation. We first consider the requirement that a transducer can be implemented
by a program using a \emph{bounded memory} (BM),
\emph{i.e.} computing the output
word using a memory independent of the size of the input word.
However when dealing with nested words in a streaming setting,
the bounded memory requirement is quite restrictive.
Indeed, even performing such a basic task as checking that a word is well-nested or
checking that a nested word belongs to a regular language of nested words
requires a memory dependent on the height (the level of nesting) of the input word~\cite{DBLP:conf/icdt/SegoufinS07}.
This observation leads us to the second question: decide, given a transducer,
whether the transduction can be evaluated with a memory that depends only
on the size of the transducer and the height of the word (but not on
its length). In that case, we say that the transduction is \emph{height bounded memory}
(HBM).  This is particularly relevant to XML transformations as
XML documents can be very long but have usually a small depth~\cite{springerlink:10.1007/s11280-005-1544-y}.
HBM does not specify \emph{how} memory depends on the height.
A stronger requirement is thus to consider HBM transductions whose
evaluation can be done with a memory that depends \emph{polynomially}
on the height of the input word.

\subsection*{Contributions}
First, we give a general space-efficient evaluation algorithm for
functional VPTs. After reading a prefix of an input word, the number
of configurations of the (non-deterministic) transducer as well as the
number of output candidates to be kept in memory may be exponential in
the size of the transducer and the height of the input word (but not
in its length). Our algorithm produces as output the longest common
prefix of all output candidates, and relies on a compact
representation of sets of configurations and remaining output
candidates (the original output word without the longest common
prefix). We prove that it uses a memory
linear in the height of the input word, and linear in the
maximal length of remaining output candidates.

We prove that BM is equivalent to sequentializability for
finite state transducers (\fsts), which is known to be decidable in
\ptime. BM is however undecidable for arbitrary pushdown transducers
but we show that it is decidable for \vpts in \conptime.

Like BM, HBM is undecidable for arbitrary
pushdown transductions. We show, via a non-trivial reduction to the emptiness
of pushdown automata with bounded reversal counters, that it is decidable in \conptime for transductions
defined by \vpts.  In particular, we show that the previously defined
algorithm runs in HBM iff the \vpt satisfies some
property, which is an extension of the so called
\emph{twinning property} for \fsts~\cite{DBLP:journals/tcs/Choffrut77}
to nested words. We call it the \emph{horizontal twinning
  property}, as it only cares about configurations of the transducers
with stack contents of identical height. This property only depends on
the transduction, \emph{i.e.} is preserved by equivalent transducers.

\manu{J'ai mis a jour le paragraphe suivant}
\olivier{Ok. J'ai juste ajout\'e $c$=call et $r$=return entre parentheses.}

When a \vpt-transduction is height bounded memory,
the memory needed may be exponential in the height of the
word. We introduce a stronger notion of height bounded memory, called
\emph{online bounded memory} (OBM). Roughly, an algorithm is OBM if the amount
of memory it uses after reading a prefix $u$ of the input nested word
only depends on the ``current'' height of $u$, i.e., the height of the
stack a visibly pushdown machine would be in after reading $u$. For
instance, $ccrrc$ has current height $1$ but height $2$
(where $c$ is a call symbol and $r$ a return symbol). We refine the
horizontal twinning property into a so called \emph{matched twinning
  property}, which we prove to effectively characterize the class of all \vpts
which can be evaluated in OBM, and to be decidable in \conptime. We call such class the class of
\emph{twinned \vpts}. We prove that twinned \vpts only require a quadratic (in the current height)
amount of memory to be evaluated. It is simple to see that any
sequentializable \vpt can be evaluated in OBM (and thus is
twinned). However, we show that some non-sequentializable \vpt are
twinned, in a way making twinned \vpt the right class of \vpt when it
comes to efficient streaming evaluation. Let us mention that the
decidability status of the class of sequentializable \vpt is open.


\subsection*{Related Work}
In the XML context, visibly pushdown automata based
streaming processing has been extensively studied for
validating XML streams~\cite{KumarMadhusudanViswanathan07,DBLP:conf/stacs/BaranyLS06,DBLP:conf/icdt/SegoufinS07}.
The  validation problem with bounded memory is studied in~\cite{DBLP:conf/stacs/BaranyLS06} when the input is assumed to be a
well-nested word and in~\cite{DBLP:conf/icdt/SegoufinS07} when it is assumed
to be a well-formed XML document (this problem is still open).
Querying XML streams has been considered in~\cite{GroheKochSchweikardt07}.
It consists in selecting a set of tuples of nodes in the
tree representation of the XML document.
For monadic queries (selecting nodes instead of tuples),
this can be achieved by a functional \vpt returning the input stream of tags,
annotated with Booleans indicating selection by the query.
However, functional \vpts cannot encode queries of arbitrary arities.
The setting for functional \vpts is in fact different to query evaluation,
because the output has to be produced on-the-fly in the right order,
while query evaluation algorithms can output nodes in any order:
an incoming input symbol can be immediately output, while
another candidate is still to be confirmed.
This makes a difference with the notion of concurrency of queries,
measuring the minimal amount of candidates to be stored,
and for which algorithms 
and lower bounds 
have been proposed~\cite{BaryossefFontouraJosifovski05}.
\vpts also relate to tree transducers~\cite{vpts10},
for which no comparable work on memory requirements is known.
However, the height of the input word is known to be a lower bound
for Core XPath filters~\cite{GroheKochSchweikardt07}.
As \vpts can express them, this lower bound also applies when evaluating \vpts.
When allowing two-way access on the input stream,
space-efficient algorithms for XML validation~\cite{KonradMagniez13}
and querying~\cite{MadhusudanViswanathan09} have been
proposed. \newcontent{Approximate space-efficient streaming
  validation algorithms of nested word properties, given as
  visibly pushdown automata, have been considered in~\cite{franois_et_al}. Finally, another related problem is the sliding
window validation problem~\cite{DBLP:conf/mfcs/GanardiJL18,DBLP:conf/icalp/GanardiHL18}: in this context, a window scans the input
and each window must satisfy some property, and the goal is to use as little memory
as possible.
}

An approach based on weighted automata for the analysis of online
algorithms has been proposed in~\cite{AKL10}. In this work, the
existence of online algorithms is related to determinism and
look-ahead removal. The analysis boils down to checking, given a
weighted automaton, whether it can be determinized or approximatively
determinized into some automaton, homomorphically embeddable into the
original one. While this problem could be adapted in our context and
is an interesting question, we did not take determinization as the
yardstick notion of streamability because, as we show, for programs
transforming nested words, deterministic \vpts are too restrictive to
capture all streamable \vpt transformations.

\manu{J'ai ajoute le paragraphe suivant. Olivier, tu avais fait une
  liste, je suis pas sur qu'il faille en parler.

\begin{itemize}
  \item introduced OBM and proved it equivalent to MTP (major difference)
  \item fixed minor bugs in algorithms
  \item harmonized decidability proofs for HTP and MTP
  \item fixed the proof that HTP implies HBM (Lemma~\ref{lm:HTP-delay})
  \item harmonized proofs HTP-implies-HBM and MTP-implies-OBM,
    and introduced arities to simplify proofs.
  \end{itemize}
}
\olivier{Ok pour moi. La liste exhaustive c'est aussi pour nous y retrouver,
  on peut la laisser en commentaires.}

\subsection*{Differences with conference version} This version
improves the results of the conference version~\cite{FiliotGauwinReynier11} both by proving stronger results, and by
simplifying proofs. Perhaps the strongest improvement is the
introduction of the class OBM, characterized by the matched twinning
property (MTP).  The MTP was already introduced in~\cite{FiliotGauwinReynier11}, but it was only shown to be a sufficient condition
for a \vpt to admit polynomially height bounded evaluation. The main
technical result of~\cite{FiliotGauwinReynier11}, based on heavy
arguments of word combinatorics, was to show that the
MTP satisfaction is invariant under equivalent \vpt, making MTP a
proper class of transductions rather than just a class of transducers. In this
journal version, we show that this class of transductions corresponds
to the class OBM, giving a full characterization in terms of memory
requirements. The word combinatorics arguments have been greatly
simplified, thanks to a recent result by Saarela~\cite{Saarela15}
about systems of word equations. The proof of Saarela's result is done in a
very elegant  way that even completely avoids word combinatorics, by
embedding words into polynomials.


\section{Visibly Pushdown Languages and Transductions}


\subsection*{Words and nested words} %
We consider
a finite alphabet $\Sigma$
partitioned into three disjoint sets $\Sigma_c$, $\Sigma_r$ and
$\Sigma_\iota$, denoting respectively the \emph{call},
\emph{return} and \emph{internal} alphabets.
We denote by $\Sigma^*$ the set of (finite) words over $\Sigma$ and by
$\epsilon$ the empty word. The length of a word $u$ is denoted by
$|u|$. \newcontent{Given $1\leq i \leq |u|$, $u[i]$ denotes the $i$-th letter of $u$}.
For all words $u,v\in\Sigma^*$, we denote by $u\wedge v$ the
longest common prefix of $u$ and $v$. More generally, for any
non-empty finite set of words $V\subseteq \Sigma^*$, the longest
common prefix of $V$, denoted by $\lcp V$, is inductively defined by
$\lcp{\{u\}} = u$ and $\lcp{V\cup\{u\}} = \lcp V \wedge u$.
We call $v$ a \emph{factor} of $u$ whenever there exist words
$v'$ and $v''$ such that $u=v'vv''$.
%
The set of \emph{well-nested} words $\Sigma^*_\wn$ is the smallest
subset of $\Sigma^*$ such that $\Sigma_\iota^*\subseteq\Sigma^*_\wn$
and for all $c\in\Sigma_c$, all $r\in \Sigma_r$, all $u,v\in
\Sigma^*_\wn$, $cur\in \Sigma^*_\wn$ and $uv\in\Sigma^*_\wn$.
Let $u=\alpha_1\dots\alpha_n\in\Sigma^*$ be a prefix of a well-nested word.
We define the \emph{current height} of $u$ as the number of pending calls:
$\hc(u)=0$ if $u$ is well-nested, and
$\hc(ucv)=\hc(u)+1$ if $c\in \Sigma_c$ and $v$ is well-nested.
The \emph{height} of $u$
is the maximal number of pending calls on any prefix of $u$, \ie,
$ \h(u) = \text{max}_{1\leq i \leq n} \hc(\alpha_1\dots\alpha_i) $.
For instance, if $c$ is a call and $r$ a return symbol, then we have $\h(crcrcc) = \h(ccrcrr) = 2$,
while $\hc(crcrcc) = 2$ and $\hc(ccrcrr)=0$. In particular, for
well-nested words, the height corresponds to the usual height of the
nesting structure of the word.

Given two words $u,v\in\Sigma^*$, the \emph{delay} of $u$ and $v$,
denoted by $\del(u,v)$, is the unique pair of
words $(u',v')$ such that  $u=(u\wedge v)u'$ and $v=(u\wedge v)v'$.
For instance, $\del(abc,abde) =
(c,de)$. Informally, in a word transduction, if there are two output
candidates $u$ and $v$ during the evaluation, we are sure that we can
output $u\wedge v$ and $\del(u,v)$ is the remaining suffixes we still
keep in memory.
We extend the concatenation to pairs of words and denote it by
$\cdot$, i.e. $(u,v)\cdot (u',v') = (uu',vv')$.
We will use the following property of delays (Lemma 5 in~\cite{Beal03}).

\begin{lem}\label{lem:assocdelay}
  For all $u,u',v,v'\in\Sigma^*$,
  $\del(uu',vv') = \del(\del(u,v)\cdot (u',v'))$.
\end{lem}


A \emph{transduction} is a binary relation $R\subseteq
\Sigma^*\times \Sigma^*$.
For any input word $u\in\Sigma^*$, we denote by $R(u)$ the
set $\{v\ |\ (u,v)\in R\}$.  A transduction $R$ is
\emph{functional} if for all $u\in\Sigma^*$, $R(u)$ has size at most
one.  If $R$ is functional, we identify $R(u)$
with the unique image of $u$ if it exists.

\subsection*{Visibly pushdown transducers (\vpts)}
As finite-state transducers
extend finite-state automata with outputs, visibly pushdown
transducers extend visibly pushdown automata~\cite{AlurJACM09}
with outputs~\cite{vpts10}. To simplify notations,
we suppose that the output alphabet is $\Sigma$, but our results
still hold for an arbitrary output alphabet. Informally, the stack
behavior of a \vpt is similar to that of visibly
pushdown automata. On a call symbol, the \vpt pushes a symbol
on the stack and produces some output word (possibly empty), on a return
symbol, it must pop the top symbol of the stack and produce some
output word (possibly empty) and on an internal symbol, the stack
remains unchanged and it produces some output word.
We do not require the output of a \vpt to be well-nested.
This is not a restriction but a more general setting as well
nestedness in the output can be enforced on the \vpt.
However, this more general setting comes for free
 as our proofs would be the same.

\begin{defi}
A \emph{visibly pushdown transducer} (\vpt) on finite words over
$\Sigma$ is a tuple $T=(Q,I, F, \Gamma, \delta)$ where
$Q$ is a finite set of states, $I\subseteq Q$ is the set of initial
states, $F\subseteq Q$ the set of final states, $\Gamma$ is the stack alphabet, $\delta=\delta_c \uplus
\delta_r\uplus \delta_\iota$ the (finite) transition relation, with
$\delta_c \subseteq Q \times\Sigma_c \times \Sigma^* \times
\Gamma \times Q$,
$\delta_r \subseteq Q \times\Sigma_r \times \Sigma^* \times
\Gamma\times Q$, and
$\delta_\iota\subseteq Q\times \Sigma_\iota\times \Sigma^* \times Q$.
\end{defi}

A \emph{configuration} of a \vpt is a pair $(q,\sigma)\in Q\times
\Gamma^*$.  A \emph{run} of $T$ on a word $u = a_1\dots
a_{l}\in\Sigma^*$ from a configuration $(q,\sigma)$ to a configuration
$(q',\sigma')$ is a finite sequence $\rho = {\{(q_k,\sigma_k)\}}_{0\leq
  k \leq l}$ such that $q_0=q$, $\sigma_0=\sigma$, $q_l=q'$,
$\sigma_l=\sigma'$ and for each $1\leq k\leq l$, there exist
$v_{k}\in\Sigma^*$ and $\gamma_k\in\Gamma$ such that either
$(q_{k-1},a_{k},v_{k},\gamma_k,q_{k})\in \delta_c$ and $\sigma_k =
\sigma_{k-1}\gamma_k$ or $(q_{k-1},a_{k},v_{k},\gamma_k,q_{k})\in
\delta_r$ and $\sigma_{k-1} = \sigma_k\gamma_k$, or
$(q_{k-1},a_k,v_k,q_k)\in\delta_\iota$ and $\sigma_k = \sigma_{k-1}$.
The word $v = v_1\dots v_{l}$ is called an \emph{output} of
$\rho$. We write $(q,\sigma)\xrightarrow{u/v} (q',\sigma')$ when there
exists a run on $u$ from $(q,\sigma)$ to $(q',\sigma')$ producing $v$
as output. We denote by $\bot$ the empty word on $\Gamma$.  A
configuration $(q,\sigma)$ is \emph{accessible} (resp.\ is
\emph{co-accessible}) if there exist $u,v\in\Sigma^*$ and $q_0\in I$
(resp.\ $q_f\in F$) such that $(q_0,\bot)\xrightarrow{u/v} (q,\sigma)$
(resp.\ such that $(q,\sigma)\xrightarrow{u/v} (q_f,\bot)$).


A transducer $T$ defines a transduction
\[
    \inter{T} = \{ (u,v)\in\Sigma^*\times\Sigma^*\ |\ \exists q\in I,q'\in F,\
(q,\bot)\xrightarrow{u/v}(q',\bot)\}.
\]
We say that a transduction $R$ is
a \vpt-transduction if there exists a \vpt $T$ such that $R =
\inter{T}$. We denote by
$T(u)$ the set $\inter{T}(u)$.

\newcontent{Two transducers $T_1,T_2$ are said to be \emph{equivalent}
  if $\inter{T_1} = \inter{T_2}$.} A transducer $T$ is \emph{reduced} if every accessible configuration
is co-accessible.
Given any \vpt, computing an equivalent reduced \vpt can be performed
in polynomial time~\cite{Caralp201513}.
A \vpt $T$ is functional if $\inter{T}$ is functional,
and this can be decided in \ptime~\cite{vpts10}.
The class of functional \vpts is denoted by \fvpt.
%
%
%
The \emph{domain} of $T$ (denoted by $\dom(T)$) is the domain of
$\inter{T}$.
The domain of $T$ contains only well-nested
words, which is not necessarily the case of the codomain.


\begin{figure}[t]
\centering
\begin{tikzpicture}[->,>=stealth',shorten >=1pt,auto,node distance=2.1cm,
                    semithick]

\node[state,accepting] (p3) {$p_3$};
\node[state]           (p2) [right of=p3] {$p_2$};
\node[state]           (p1) [right of=p2] {$p_1$};
\node[state,initial above]   (i)  [right of=p1] {$i$};
\node[state]           (q1) [right of=i]  {$q_1$};
\node[state]           (q2) [right of=q1] {$q_2$};
\node[state,accepting] (q3) [right of=q2] {$q_3$};


\path (i)  edge [above]      node {$c/a, \gamma$} (p1)
      (p1) edge [loop above] node {$c/a, \gamma$} (p1)
           edge [above]      node {$r/c, \gamma$} (p2)
      (p2) edge [loop above] node {$r/c, \gamma$} (p2)
           edge [above]      node {$r/c, \gamma$} (p3);

\path (i)  edge [above]      node {$c/b, \gamma$} (q1)
      (q1) edge [loop above] node {$c/b, \gamma$} (q1)
           edge [above]      node {$r/c, \gamma$} (q2)
      (q2) edge [loop above] node {$r/c, \gamma$} (q2)
           edge [above]      node {$r'/c, \gamma$} (q3);


\path (q2) edge [above,dashed,bend right=45] node {$c/b, \gamma$} (q1);
\path (p2) edge [above,dashed,bend left=45]  node {$c/a, \gamma$} (p1);

\end{tikzpicture}
\caption{A functional \vpt with $\Sigma_c=\{c\}$, $\Sigma_r=\{r,r'\}$
  and $\Sigma_\iota = \{a,b\}$}\label{fig:vpt3}
\end{figure}
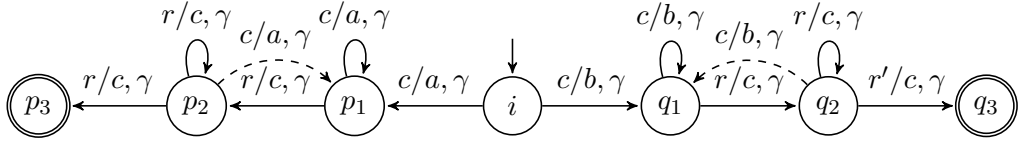


\begin{exa}\label{ex:vpt1}
  Consider the \vpt $T$ of \Figure~\ref{fig:vpt3} represented in plain
  arrows. The left and right parts accept the same input words except
  for the last letter of the word. The domain of $T$ is $\dom(T) = \{
  c^{n}r^{n}\ |\ n\geq 2\}\cup \{ cc^{n}r^{n}r'\ |\ n\geq 1\}$. Any word
  $c^{n}r^{n}$ is translated into $a^{n}c^{n}$, and any word $cc^{n}r^{n}r'$ is
  translated into $b^{n+1}c^{n+1}$. Therefore the translation of the
  first sequence of calls depends on the last letter $r$ or $r'$.  As we will see later,
  this transduction cannot be evaluated with a bounded amount of memory,
  but with a memory which depends on the height $n$ of the input word.
\end{exa}

\subsection*{Finite state transducers (\fsts)} %
A \emph{finite state transducer} (\fst) on an alphabet $\Sigma$ is
a tuple $(Q,I,F,\delta)$
where $Q$ is a finite set,
$I,F\subseteq Q$ and $\delta\subseteq Q\times \Sigma\times
\Sigma^*\times Q$ with the standard semantics.
This definition corresponds to the usual definition of
\emph{real-time} \fsts~\cite{sakarovich:2009a}, as there is no $\epsilon$-transitions. We
always consider real-time \fsts in this paper, so we just call them \fsts.

\subsection*{\newcontent{Sequential transducers}}
The \emph{underlying automaton} of a given \fst (resp. \vpt) is the automaton obtained by ignoring
the output. A \emph{sequential} \fst (resp. \vpt) is a pair $(T,\Psi)$ where
$T$ is an \fst (resp. \vpt) whose underlying automaton is deterministic, and
$\Psi:F\rightarrow \Sigma^*$ is a mapping that associates a word with
each final state. The output of an input word $u$ by $(T,\Psi)$ is the
word $v.\Psi(q)$ if the unique run of $T$ on $u$
  produces $v$ and ends in some accepting state $q$.

\subsection*{\newcontent{\vpts on words of bounded height}}
Given a natural number $k\in \Nat$ and a \vpt $T$,
one can define an \fst, denoted by
$\fst(T,k)$, which is the restriction of $T$ to input words of height
less than $k$. The transducer \newcontent{$\fst(T,k)$} is naturally constructed by taking as
states the configurations $(q,\sigma)$ of $T$ such that $|\sigma|\leq
k$. \newcontent{In particular, its initial (resp.\ final) states are the pairs
  $(q,\bot)$ where $q$ is initial (resp.\ final), and there is a
  transition in $\fst(T,k)$ from state $(q,\sigma)$ to state $(p,\gamma)$ on reading
  $u\in\Sigma$, producing $v\in\Sigma^*$, if there is a (single)
  transition in $T$
  from the configuration $(q,\sigma)$ to the configuration
  $(p,\gamma)$ on input $u$ and output $v$.}

\subsection*{Turing Transducers}
In order to formally define the complexity classes for evaluation that we target,
we introduce a \emph{deterministic} computational model for word
transductions that we call \emph{Turing Transducers}. Turing
transducers, \newcontent{a special case of Turing machines}, have
three tapes: one read-only left-to-right input tape over some alphabet
$\Sigma$, one write-only
left-to-right output tape over $\Sigma$, and one standard working
tape over some alphabet $\Sigma'$. \newcontent{Their transitions are
 assumed to be deterministic, to model deterministic algorithms.}
They have accepting and rejecting
states. A functional transduction $f : \Sigma^*\rightarrow \Sigma^*$ is
\emph{computable} by a Turing transducer $T$ if for all $u\in \dom(f)$,
the machine $T$ halts in some accepting state and the content of the
output tape is $f(u)$, and for all $u\not\in\dom(f)$, the machine
halts in some rejecting state. The space complexity of a Turing
transducer is measured on the working tape only.



\section{Bounded Memory Evaluation}%
\label{sec:pb}\label{sec:boundedmem}

In this section, we consider the class of transductions that can be
evaluated with a constant amount of memory if we fix the machine that
defines the transduction, and the problem of deciding whether a
transducer (finite-state, pushdown, or visibly pushdown) defines a
transduction in this class.

\begin{defi}
A (functional) transduction $f : \Sigma^*\rightarrow \Sigma^*$ is
\emph{bounded memory (BM)} if there exists $K\in\Nat$ such that
it is computable by a Turing transducer $M$ that runs in
space complexity at most $K$.
\end{defi}

\begin{exa}
    Let $\Sigma = \{a,b\}$ be an alphabet and let $f$ be the
    transduction that maps any word of the form $w\sigma$, for
    $\sigma\in\Sigma$ and $w\in\Sigma^*$, to $\sigma w$. Clearly, $f$
    is not BM\@: any Turing transducer that computes this transduction,
    since it reads the input from left-to-right, and produces the
    output from left-to-right, must wait until the last letter of the
    word before outputting anything, and therefore has to store on the
    working tape the word $w$.

    \newcontent{As a positive example, any function $f$ on a finite domain $D$ is BM, by taking
      $K$ as the length of the longest output word of $f$ on $D$. The
      domain $D$ needs not to be finite in general for $f$ to be
      BM\@. Indeed, as we show, all sequential functions are BM, and conversely.}
\end{exa}

\subsection{Finite state transducers} It is not difficult to verify that for
\fst-transductions, bounded memory is characterized by
sequentializability, which is decidable in \ptime. An \fst $T$ is
sequentializable if there exists a sequential transducer $T_d$
such that $\inter{T} = \inter{T_d}$. Sequentializable transducers
have been characterized by a structural property of their runs, called
the twinning property~\cite{DBLP:journals/tcs/Choffrut77},
which is decidable in \textsf{PTime}~\cite{DBLP:journals/iandc/WeberK95,Beal03}.
Intuitively, this property requires that no delay can be accumulated
along loops synchronized on the same input.

\begin{defi}[Twinning property for \fsts]\label{def:tp2}
Let $T = (Q, I, F, \delta)$ be a reduced \fst. $T$ satisfies the twinning property if
for all $q_0,q'_0\in I$, for all $q,q'\in Q$,  for all words $u_1,v_1,w_1,u_2,v_2,w_2\in\Sigma^*$, if:
\[
q_0\ \xrightarrow{u_1/v_1}\ q\ \xrightarrow{u_2/v_2}\ q\qquad\qquad
q'_0\ \xrightarrow{u_1/w_1}\ q'\ \xrightarrow{u_2/w_2}\ q'
\]
then $\Delta(v_1,w_1) = \Delta(v_1v_2,w_1w_2)$.
\end{defi}

\begin{prop}\label{prop:bmfst}
  Let $T$ be a functional \fst. The following statements are equivalent:
  \begin{enumerate}
  \item $\inter{T}$ is BM
  \item $T$ is sequentializable
  \item $T$ satisfies the twinning property.
  \end{enumerate}

  Moreover, it is decidable in \ptime whether $\inter{T}$ is BM\@.
\end{prop}

\begin{proof}
  The equivalence between $(2)$ and $(3)$ has been shown in~\cite{DBLP:journals/tcs/Choffrut77}.
  We show the equivalence between~$(1)$ and~$(2)$. Clearly, if $\inter{T}$ is definable by a sequential transducer
  $T_d$, then evaluating $T_d$ on any input word $u$ can be done with
  a space complexity that depends on the size of $T_d$ only.

  Conversely, if $\inter{T}$ is BM, there exists $K\in\Nat$ and a Turing
  transducer $M$ that transforms any input word $u$ into $\inter{T}(u)$ in space
  complexity $K$. Any word on the working tape of $M$ is of length at
  most $K$. As $M$ is deterministic, we can therefore see $M$ as a
  sequential \fst, whose states are pairs $(q,w)$ where $q$ is a
  state of $T$ and $w$ a word on the working tape (modulo some
  elimination of $\epsilon$-transitions).

  Since sequentializability is decidable in \ptime, as first shown
  in~\cite{DBLP:journals/iandc/WeberK95}, and later on with a
  different proof in~\cite{Beal03}, the result follows from
  the equivalence between~$(1)$ and~$(2)$.
\end{proof}

\subsection{Pushdown transducers} Similarly to finite-state
transducers that extend finite-state automata with outputs, pushdown
transducers extend pushdown automata with outputs. Bounded Memory is undecidable for
pushdown transducers, since it is at least as difficult as deciding whether a
\newcontent{(non-deterministic)} pushdown automaton defines a regular
language \newcontent{(the reduction is immediate)}.

\begin{prop}\label{prop:pt}
  It is undecidable whether a functional transduction defined by a
  (non-deterministic) pushdown transducer is BM\@.
\end{prop}



Prop.\ref{prop:pt} holds for non-deterministic pushdown
transducers. It is open whether it holds too for deterministic pushdown
transducers. The same reduction cannot be applied since testing the
regularity of the language defined by a deterministic pushdown
automaton is decidable~\cite{Stearns67}.

\subsection{Visibly pushdown transducers}

For \vpts, BM is quite restrictive as
it imposes to verify
whether a word is well-nested by using a bounded amount of
memory. This can be done only if the height of the words of the domain
is bounded by some constant which depends on the transducer only:
\begin{prop}\label{thm:bmvpts}
  Let $T$ be a functional \vpt with $n$ states.
  \begin{enumerate}
    \item $\inter{T}$ is BM iff $(i)$ for all $u\in\dom(T)$, $\h(u)\leq n^2$, and
      $(ii)$ $\inter{\fst(T,n^2)}$ is BM\@;
    \item It is decidable in \conptime whether $\inter{T}$ is BM\@.
  \end{enumerate}
\end{prop}

\begin{proof}
  If $\inter{T}$ is BM, there exist $K$ and a Turing transducer $M$
  computing $\inter{T}$, and such that
  $M$ evaluates any input word in space at most $K$. We can easily extract
  from $M$ a finite automaton that defines $\dom(T)$, whose number of
  states $m$ only depends on $M$ and $K$. By a simple pumping
  argument, it is easy to show that the words in $\dom(T)$ have a
  height bounded by $m$. If the height of the words in $\dom(T)$ is
  bounded, then it is bounded by $n^2$. Indeed, assume that
  there exists a word $u\in\dom(T)$ whose height is strictly larger
  than $n^2$.
  Consider all decompositions of $u$ into nested well-nested factors,
  \ie, $u=u_1u_2u_3u_4u_5$ where $u_3$ and $u_2u_3u_4$ are well-nested,
  and $\hc(u_2)>0$.
  As the height of $u$ is strictly larger than $n^2$, there exists one of
  such decompositions for which the states $q,p$
  reached respectively before $u_2$ and after $u_4$ will repeat around $u_3$.
  In other words at least one run of $T$ on $u$ has the following form:
  \[
  (i,\bot) \xrightarrow{u_1/v_1} (q,\sigma) \xrightarrow{u_2/v_2}
  (q,\sigma\sigma') \xrightarrow{u_3/v_3} (p,\sigma\sigma')
  \xrightarrow{u_4/v_4} (p,\sigma) \xrightarrow{u_5/v_5} (f,\bot)
  \]
  with $\sigma'$ non-empty, and $i$
  (resp. $f$) an initial (resp.\ final) state of $T$.
  Then one can iterate the
  matching loops around $q$ and $p$ to generate words
  $u_{1}u_{2}^{k}u_{3}u_{4}^{k}u_{5}$ in $\dom(T)$
  with arbitrarily large heights, yielding a contradiction.
  Therefore $\fst(T,n^2)$ is equivalent to $T$. As in the proof of
  Proposition~\ref{prop:bmfst}, we can consider $M$ as a sequential \fst
  $T_M$ whose set of states are configurations of the machine. The
  \fst $T_M$ is equivalent to $T$, and therefore to
  $\fst(T,n^2)$. Since $T_M$ is sequential, $\fst(T,n^2)$ is
  sequentializable and therefore by Proposition~\ref{prop:bmfst},
  $\inter{\fst(T,n^2)}$ is BM\@. The converse is obvious.

  Therefore to check whether $\inter{T}$ is BM, we first decide if the height of all input words
  accepted by $T$ is less or equal than $n^2$. This can be done in
  \ptime $O(|T|\cdot n^2)$ by checking emptiness of the projection of
  $T$ on the inputs (this is a visibly pushdown automaton) extended
  with counters up to $n^2+1$ that count the height of the word.
  \newcontent{One can then construct $\fst(T,n^2)$, resulting in an
    exponentially larger $\fst$ equivalent to $T$, and check whether
    $\inter{\fst(T,n^2)}$ is BM using the procedure of
  Theorem~\ref{prop:bmfst}. The time complexity of the overall
  algorithm is exponential. However, using results which are proved
  later to characterise a more general class of transductions (namely the
  \vpt-transductions which can be evaluated with height bounded
  memory, forming the class called HBM --- Definition~\ref{def:hbm}),
  one can lower the complexity to \conptime. By definition of HBM, a
  \vpt-transduction whose input words have bounded height (i.e., a
  height which only depends on the transducer itself) is HBM iff
  it is BM\@. It is shown in Theorem~\ref{thm:vptsboundedmem} that HBM
  can be tested in \conptime, yielding the result.
\newline
  For the sake of
  completeness, let us give the main arguments to get the \conptime
  bound. We use pushdown counter machines which make a bounded number
  of reversals (a bounded number of moves from an increasing to a
  decreasing mode, and from a decreasing to an increasing mode). Such
  machines are known to have decidable emptiness problem in
  \conptime~\cite{vpts10}. This counter machine accepts nested words
  on which there are two runs witnessing the non-satisfiability of the
  twinning property by $\fst(T,n^2)$. It is not necessary to encode
  the stack explicitly in the state, using the pushdown mechanism of
  the counter machine, and hence we can keep the size of the counter
  machine polynomial (in $T$). To witness the non-satisfiability of
  the twinning property, one uses combinatorics properties of the
  output words produced by those runs in case the delays are
  different. Their are several conditions to be checked (taken in disjunction), one of them
  being that there is a mismatch between the output $v_1$ and the
  output of $v_2$, i.e., there is a position $i$ such that $v_1[i]\neq
  v_2[i]$. The counter machine simulates the behaviour of $T$
  (without producing anything) and the difficulty is that the $i$-th
  position of $v_1$ and $v_2$ may not produced when reading
  \emph{different} input positions. The machine instead
  non-deterministically guesses to output positions $c_1,c_2$ (whose
  values are stored in two different counters), check that
  $v_1[c_1]\neq v_2[c_2]$ and later on checks that $c_1=c_2$. This can
  be done using only one reversal. The details can be found in the
  proof of Proposition~\ref{prop:HTP}.
}
\end{proof}

Note that in order to decide whether a functional \vpt $T$ with $n$
states defines a transduction in BM, one could proceed as follows: first decide whether
all the nested words of the domain have height at most $n^2$, then
construct $\fst(T,n^2)$, and then decide whether $\fst(T,n^2)$ is
sequentializable using Prop.\ref{prop:bmfst}. This would however
gives an \textsf{ExpTime} procedure, as $\fst(T,n^2)$ has exponential
size, since there are exponentially many stack contents of height
$n^2$ in general.


\section{Online Evaluation Algorithm of \vpt-Transductions}%
\label{sec:algo}

We present an online algorithm \eval to evaluate functional word transductions
defined by \fvpts\footnote{\newcontent{We remind the reader that \fvpts stand for the
class of functional \vpts.}}.
For clarity, we present this algorithm under some assumptions,
without loss of generality.
First, input words of our algorithms
are words $u\in\Sigma^*$ concatenated with a special symbol $\EOS\ \notin\Sigma$,
denoting the end of the word.
Second, we only consider input words without internal symbols
($\Sigma_\iota = \varnothing$), as they can easily be encoded by successive call and return symbols.
Third, we assume an implementation of \vpts such that
the set \newcontent{$H$} of transitions with a given left-hand side can be
retrieved in time $O(|H|)$.

\begin{algorithm}[t]
  \begin{algorithmic}[2]

    \Procedure{\eval}{\fvpt $T$, function $\nextChar()$}
      \State reduce($T$)   \COMMENT[.75\linewidth]{in PTIME using~\cite{Caralp201513}} 
      \State initialize($S$)   \COMMENT[.75\linewidth]{DAG with edges $\rootstr\stredgesymb{\emptyword}{(q_0,\bot,0)}$ for all initial states $q_0$ of $T$} 
      \State $\alpha \leftarrow \nextChar()$   \COMMENT[.75\linewidth]{read first input symbol}
      \While {$\alpha \neq \EOS$}
        \If {$\alpha$ is a return symbol}
          \If {$S$.height() $\leq 1$}   \COMMENT{pop on empty stack} 
            \State reject this input word
          \Else
            \State $\text{update\_return}(S,T,\alpha)$   \COMMENT{see Algorithm~\ref{algo:update-str-return}}
          \EndIf
        \Else    \COMMENT{$\alpha$ is a call symbol}
            \State $\text{update\_call}(S,T,\alpha)$   \COMMENT{see Algorithm~\ref{algo:update-str-call}}
        \EndIf
        \State $\text{output\_lcp}(S)$   \COMMENT{see Algorithm~\ref{algo:outputlcp}}
        \State $\alpha \leftarrow \nextChar()$   \COMMENT{read next input symbol}
      \EndWhile
      \If {$S$.height()$=1$ and $\rootstr\stredgesymb{v}{(q,\bot,0)}$ in $S$ with $q$ final state of $T$} 
        \State \textbf{output} $v$
        \State accept this input word
      \Else
        \State reject this input word
      \EndIf
    \EndProcedure
  \end{algorithmic}
  \caption{\label{algo:lcpin} \quad
    Algorithm \eval.}
\end{algorithm}

\olivier{Algorithm~\ref{algo:lcpin} ajout\'e, ainsi que sa presentation ci-dessous.}
The core task of this algorithm, presented in Algorithm~\ref{algo:lcpin}, is to maintain
the configuration for each run of the \fvpt $T$ on the input $u$,
and produce its output on-the-fly.
These configurations are efficiently stored in a data structure $S$.
The first step of the algorithm \eval is to transform $T$
into a reduced \fvpt in polynomial time, using~\cite{Caralp201513}.
Indeed, when $T$ is reduced, functionality ensures that,
for a given input word $u$,
and for every accessible configuration $(q,\sigma)$ of $T$,
there is at most one $v$ such that
$(q_i,\bot)\xrightarrow{u/v}{(q,\sigma)}$ with $q_i$ an initial state.
Hence, we define a notion called  \emph{\dconf}, as  triples $(q,\sigma,w)$,
where $q$ is the current state of the run, $\sigma$ its corresponding stack
content, and $w$ is a suffix of $v$, which has not been output yet.

The set $C$ of \dconfs of $T$ on an input word $u$
can be incrementally computed,
starting from the set $\{ (q_i,\bot,\epsilon)\ |\ q_i\in I \}$,
and updated in the following way after reading a call symbol $c\in\Sigma_c$:
\begin{displaymath}
\update(C,c) = \bigcup_{(q,\sigma,v)\in C}
\{ (q',\sigma\gamma,vv') \hmid (q,c,v',\gamma,q')\in\delta_c \}
\end{displaymath}
and, for a return symbol $r\in\Sigma_r$:
\begin{displaymath}
\update(C,r) = \bigcup_{(q,\sigma\gamma,v)\in C}
\{ (q',\sigma,vv') \hmid (q,r,v',\gamma,q')\in\delta_r \}
\end{displaymath}

This provides us a first algorithm: update the set of \dconfs until the
last letter $\EOS$, and then output the (unique) word $v$
\newcontent{shared by all \dconfs} in this set.
This algorithm is inefficient in two aspects.
First, it explicitly stores all \dconfs, and this set may grow exponentially.
Second, it does not output anything before the end, and thus stores parts
of the output that could have been released before the end, saving memory.
Algorithm \eval addresses these two flaws thanks to the two
following features.

\subsection{Compact representation}

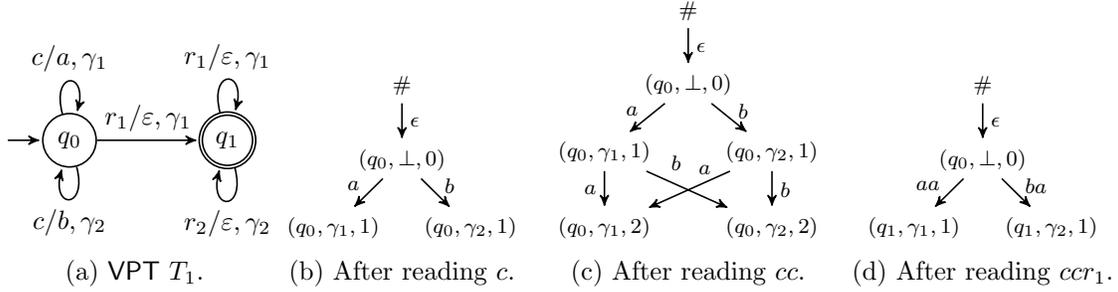
\begin{figure}[t]
\centering

\begin{tabular}{@{}c@{}ccc}

\begin{tikzpicture}[->,>=stealth',shorten >=1pt,auto,node distance=2.1cm,
    semithick]
\small
\node[state,initial]   (q0) {$q_0$};
\node[state,accepting] (q1) [right of=q0] {$q_1$};
\path (q0) edge [loop above] node {$c / a, \gamma_1$} (q0)
      (q0) edge [loop below] node {$c / b, \gamma_2$} (q0)
      (q0) edge [above]      node {$r_1 / \varepsilon, \gamma_1$} (q1)
      (q1) edge [loop above] node {$r_1 / \varepsilon, \gamma_1$} (q1)
      (q1) edge [loop below] node {$r_2 / \varepsilon, \gamma_2$} (q1);
\end{tikzpicture}

&

\begin{tikzpicture}[->,>=stealth',shorten >=1pt,auto,semithick]
\scriptsize 
\node (rootstr)                                   {$\rootstr$};
\node (q0b0) [below=.5cm of rootstr]              {$(q_0,\bot,0)$};
\node (q011) [below left =.4cm and -.5cm of q0b0] {$(q_0,\gamma_1,1)$};
\node (q021) [below right=.4cm and -.5cm of q0b0] {$(q_0,\gamma_2,1)$};
\draw (rootstr) -- node [right] {$\emptyword$} (q0b0);
\draw (q0b0) -- node [left,yshift=.1cm]  {$a$} (q011);
\draw (q0b0) -- node [right,yshift=.1cm] {$b$} (q021);
\end{tikzpicture}

&

\begin{tikzpicture}[->,>=stealth',shorten >=1pt,auto,semithick]
\scriptsize 
\node (rootstr)                                   {$\rootstr$};
\node (q0b0) [below=.5cm of rootstr]              {$(q_0,\bot,0)$};
\node (q011) [below left =.4cm and -.3cm of q0b0] {$(q_0,\gamma_1,1)$};
\node (q021) [below right=.4cm and -.3cm of q0b0] {$(q_0,\gamma_2,1)$};
\node (q012) [below of=q011]                      {$(q_0,\gamma_1,2)$};
\node (q022) [below of=q021]                      {$(q_0,\gamma_2,2)$};
\draw (rootstr) -- node [right] {$\emptyword$} (q0b0);
\draw (q0b0) -- node [left,yshift=.1cm]  {$a$} (q011);
\draw (q0b0) -- node [right,yshift=.1cm] {$b$} (q021);
\draw (q011) -- node [left]              {$a$} (q012);
\draw (q011) -- node [pos=.2]            {$b$} (q022);
\draw (q021) -- node [pos=.3,above]      {$a$} (q012);
\draw (q021) -- node                     {$b$} (q022);
\end{tikzpicture}

&

\begin{tikzpicture}[->,>=stealth',shorten >=1pt,auto,semithick]
\scriptsize 
\node (rootstr)                                   {$\rootstr$};
\node (q0b0) [below=.5cm of rootstr]              {$(q_0,\bot,0)$};
\node (q011) [below left =.4cm and -.5cm of q0b0] {$(q_1,\gamma_1,1)$};
\node (q021) [below right=.4cm and -.5cm of q0b0] {$(q_1,\gamma_2,1)$};
\draw (rootstr) -- node [right] {$\emptyword$} (q0b0);
\draw (q0b0) -- node [left,yshift=.1cm]  {$aa$} (q011);
\draw (q0b0) -- node [right,yshift=.1cm] {$ba$} (q021);
\end{tikzpicture}

\\

{\small (a) \vpt $T_1$.}
&
{\small (b) After reading~$c$.}
&
{\small (c) After reading $cc$.}
&
{\small (d) After reading $ccr_1$.}
\end{tabular}

\caption{Data structure used by \eval.%
\label{fig:eval-compact}}
\end{figure}

First, the set of current \dconfs is stored in a compact structure
that shares common stack contents.
Consider for instance the \vpt $T_1$ in \Figure~\ref{fig:eval-compact} (a).
After reading $cc$, current \dconfs are
$\{(q_0,\gamma_1\gamma_1,aa)$, $(q_0,\gamma_1\gamma_2,ab)$,
   $(q_0,\gamma_2\gamma_1,ba)$, $(q_0,\gamma_2\gamma_2,bb)\}$.
Hence after reading $c^n$, the number of current \dconfs is $2^n$.
However, \dconfs share a lot of information. For instance, the previous set is the set of tuples
$(q_0,\eta_1\eta_2,\alpha_1\alpha_2)$ where
$(\eta_i,\alpha_i)$ is either $(\gamma_1,a)$ or $(\gamma_2,b)$.

Based on this observation, we propose a data structure avoiding this blowup.
As illustrated in \Figure~\ref{fig:eval-compact}~(b) to (d),
this structure is a directed acyclic graph (DAG).
The root of this DAG is denoted by $\rootstr$, and the other nodes
are tuples $(q,\gamma,i)$ where $q\in Q$,
$\gamma\in\Gamma$ and $i\in\Nat$ is the depth of the node in the DAG\@.
Each edge of the DAG, denoted by $\stredgesymb{}$, is labelled with a word,
so that a branch of this DAG, read from the root $\rootstr$ to the leaf,
represents a \dconf $(q,\sigma,v)$:
$q$ is the state in the leaf, $\sigma$ is the concatenation
of stack symbols in traversed nodes, and $v$ is the concatenation of
words on edges.
For instance, in the DAG of \Figure~\ref{fig:eval-compact}~(c), the branch
$\rootstr\stredgesymb{\emptyword}{(q_0,\bot,0)}\stredgesymb{b}{(q_0,\gamma_2,1)}
\stredgesymb{a}{(q_0,\gamma_1,2)}$
encodes the \dconf $(q_0,\gamma_2\gamma_1,ba)$ of the \vpt of
\Figure~\ref{fig:eval-compact}~(a).
However, this data structure cannot store any set of accessible \dconfs
of arbitrary functional \vpts:
at most one delay $w$ has to be assigned to a \dconf. This is why we need $T$ to be reduced.

We denote by $\str u T$ the DAG obtained after reading $u$.
When a call letter $c\in\Sigma_c$ is read, the structure $\str u T$ is updated
such that, for every leaf of $\str u T$,
a child is added for every way of updating the
corresponding configuration according to a rule of $T$.
If a leaf cannot be updated, it is removed and this removal is propagated
upwards to its ascendants becoming leaves (procedure \textsc{remove\_edges}).
Algorithm~\ref{algo:update-str-call} describes how $\str {uc} T$ is
computed from $\str u T$.
For sake of clarity, we only show how edges are updated, not nodes
(nodes without incoming edges are automatically removed).
\begin{algorithm}[t]
  \begin{algorithmic}[2]

    \Procedure{update\_call}{structure $S$, transducer $T$, call symbol $c$}
      \State $\newEdges \leftarrow \emptyset$
      \State $\orphans \leftarrow \emptyset$
      \For {$(q,\gamma,i)\in\strleaves{S}$}
        \If {$\exists v,\gamma',q' \hmid (q,c,v,\gamma',q')\in\delta_T$}
          \For {$(v,\gamma',q') \hmid (q,c,v,\gamma',q')\in\delta_T$}
            \State $\newEdges.\add(\stredge{(q,\gamma, i)}{v}{(q',\gamma',i+1)})$
          \EndFor
        \Else
          \State $\orphans.\add((q,\gamma,i))$
        \EndIf
      \EndFor
      \State $\removeEdges(S, \orphans)$
      \State $\stredges{S}.add(\newEdges)$
    \EndProcedure
    \State

    \Procedure{remove\_edges}{structure $S$, set $\orphans$}
      \While {$\orphans\not=\emptyset$}
        \State $n\leftarrow \orphans.\pop()$
        \For {$m \hmid \exists v,\ \stredge{m}{v}{n}$}
          \State $S.\remove(\stredge{m}{v}{n})$
          \If {$\nexists n',v',\ \stredge{m}{v'}{n'}$}
            $\orphans.\add(m)$
          \EndIf
        \EndFor
      \EndWhile
    \EndProcedure

  \end{algorithmic}
  \caption{\label{algo:update-str-call} \quad
    Updating structure $S$ with a call symbol.}
\end{algorithm}
For a return letter $r\in\Sigma_r$, we try to pop every leaf:
if it is possible, the leaf is removed and the new leaves updated,
otherwise we remove the leaf and propagate the removal upwards
(procedure \textsc{remove\_edges}).
This is described in Algorithm~\ref{algo:update-str-return},
where the future level $i-1$ is stored in $newEdges$,
then levels $i$ and $i-1$ are removed by two calls to \textsc{remove\_leaves},
and finally the new level $i-1$ is added.
\begin{algorithm}[t]
  \begin{algorithmic}[2]

    \Procedure{update\_return}{structure $S$, transducer $T$, return symbol $r$}
      \State $\newEdges \leftarrow \emptyset$
      \State $\orphans \leftarrow \emptyset$
      \For {$(q_\ell,\gamma_\ell,i)\in\strleaves{S}$}
        \If {$\exists v,q \hmid (q_\ell,r,v,\gamma_\ell,q)\in\delta_T$}
          \For {$(v,q) \hmid (q_\ell,r,v,\gamma_\ell,q)\in\delta_T$}
            \For {$(q_0,\gamma_0,v_0) \hmid
              \stredge{(q_0,\gamma_0,i-1)}{v_0}{(q_\ell,\gamma_\ell,i)}
              \in\stredges{S}$}
              \For {$(n,v_1) \hmid
                \stredge{n}{v_1}{(q_0,\gamma_0,i-1)}
                \in\stredges{S}$}
                \State $\newEdges.\add(\stredge{n}{v_1v_0v}{(q,\gamma_0,i-1)})$
              \EndFor
            \EndFor
          \EndFor
        \Else
          \State $\orphans.\add((q_\ell,\gamma_\ell,i))$
        \EndIf
      \EndFor
      \State $\removeEdges(S, \orphans)$
      \State $\removeLeaves(S)$ \quad // level $i$
      \State $\removeLeaves(S)$ \quad // level $i-1$
      \State $\stredges{S}.add(\newEdges)$
    \EndProcedure
    \State

    \Procedure{remove\_leaves}{structure $S$}
      \For {$n\in\strleaves{S}$}
        \For {$(m,v) \hmid \stredge{m}{v}{n} \in \stredges{S}$}
          \State $S.\remove(\stredge{m}{v}{n})$
        \EndFor
      \EndFor
    \EndProcedure
  \end{algorithmic}
  \caption{\label{algo:update-str-return} \quad
    Updating structure $S$ with a return symbol.}
\end{algorithm}
%

The correctness of this construction can be established by proving the
following invariant by induction on $|u|$:
\begin{quote}
For every $0\le i\le \hc(u)$, there is a path labelled by $v$
from the root $\rootstr$ to the node $(q,\sigma,i)$ in $\str u T$
iff
there exists $q_0\in I$ such that
$(q_0,\bot)\xrightarrow{u_1\cdots u_k/v}{(q,\sigma)}$
where $k=\max \{j \hmid \hc(u_1\cdots u_j) = i\}$.
\end{quote}

\subsection{Computing outputs}

The second main feature of \eval is that it ensures that
after reading a prefix $u'$ of a word $u$,
it will have output the longest common prefix of all corresponding runs,
\ie, the word $\lcpin {u'} T = \lcp{\reach(u')}$ where
\[
\reach(u') =
\{v\hmid \exists (q_0,q,\sigma)\in I\times Q\times \Gamma^*,\
(q_0,\bot)\xrightarrow{u'/v}(q,\sigma)\}.
\]

As detailed in Algorithm~\ref{algo:lcpin},
when a new input symbol is read, the DAG is first updated as described
in the previous section,
using Algorithm~\ref{algo:update-str-call} for a call symbol,
and Algorithm~\ref{algo:update-str-return} for a return symbol.

\begin{figure}
\centering

\begin{tabular}{cc}

\begin{tikzpicture}[->,>=stealth',shorten >=1pt,auto,semithick]
\node (n) {$n$};
\node (i1) [above left =1cm and 1cm of n] {};
\node (i2) [above right=1cm and 1cm of n] {};
\node (o1) [below left =1cm and 1cm of n] {};
\node (o2) [below=1.1cm of n]             {};
\node (o3) [below right=1cm and 1cm of n] {};
\draw (i1) -- node                     {$a$}   (n);
\draw (i2) -- node [above,xshift=-1mm] {$b$}   (n);
\draw (n)  -- node [above,xshift=-2mm] {$abc$} (o1);
\draw (n)  -- node [pos=.6]            {$aba$} (o2);
\draw (n)  -- node [above,xshift=2mm]  {$aba$} (o3);
\end{tikzpicture}
&
\begin{tikzpicture}[->,>=stealth',shorten >=1pt,auto,semithick]
\node (n) {$n$};
\node (i1) [above left =1cm and 1cm of n] {};
\node (i2) [above right=1cm and 1cm of n] {};
\node (o1) [below left =1cm and 1cm of n] {};
\node (o2) [below=1.1cm of n]             {};
\node (o3) [below right=1cm and 1cm of n] {};
\draw (i1) -- node                     {$aab$}   (n);
\draw (i2) -- node [above,xshift=-2mm] {$bab$}   (n);
\draw (n)  -- node [above,xshift=-2mm] {$c$} (o1);
\draw (n)  -- node [pos=.6]            {$a$} (o2);
\draw (n)  -- node [above,xshift=2mm]  {$a$} (o3);
\end{tikzpicture}
\\
(a) Internal node $n$ of the DAG\@.
&
(b) Node $n$ after update by \factorize.
\\
\end{tabular}

\caption{Effect of \factorize on a node.}\label{fig:factorize}
\end{figure}
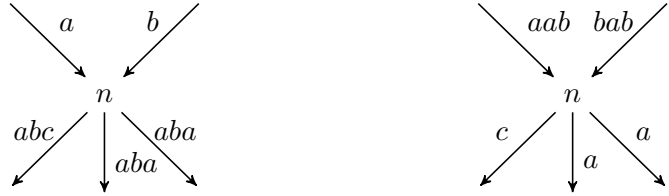

Then, a bottom-up pass on this DAG computes and outputs $\lcpin {u'} T$
as described by Algorithm~\ref{algo:outputlcp}.
This one starts with the procedure \factorize,
that processes every node in a bottom-up manner
(from leaves to the root \rootstr).
For each node (except the root), let $\ell$ be the longest common prefix
of labels of outgoing edges.
Then $\ell$ is removed from these outgoing edges, and concatenated at the end of
labels of incoming edges.
This is illustrated in \Figure~\ref{fig:factorize}.
At the end, the longest common prefix of all output words on $u'$
is the longest common prefix of the words labelling the edges outgoing
from the root node $\rootstr$.
It can be easily shown by induction on the DAG that \factorize preserves
the set of \dconfs stored in this DAG\@.

\begin{algorithm}[ht]
  \begin{algorithmic}[2]

    \Procedure{output\_lcp}{structure $S$}
      \State $\factorize(S, \rootstr, \emptyset)$
      \State $\ell \leftarrow \lcp{\{ v \hmid \exists n,\ \stredge{\rootstr}{v}{n}\}}$
      \State \textbf{output} $\ell$
      \For {$n,v \hmid \stredge{\rootstr}{\ell\cdot v}{n}$}
        \State replace $\stredge{\rootstr}{\ell\cdot v}{n}$ by $\stredge{\rootstr}{v}{n}$ in $S$
      \EndFor
    \EndProcedure
    \State

    \Function{factorize}{structure $S$, node $n$, set $\done$}
      \If {$n\notin\strleaves{S}$}
      \For {$m,v \hmid \stredge{n}{v}{m}$ and $m\notin \done$}
        \State $\done\leftarrow \factorize(S, m, \done)$
      \EndFor
      \If {$n\not=\rootstr$}
      \State $\ell\leftarrow\lcp{\{ v \hmid \exists m,\ \stredge{n}{v}{m}\}}$
      \For {$m,v \hmid \stredge{n}{\ell\cdot v}{m}$}
        \State replace $\stredge{n}{\ell\cdot v}{m}$ by $\stredge{n}{v}{m}$ in $S$
      \EndFor
      \For {$m,v \hmid \stredge{m}{v}{n}$}
        \State replace $\stredge{m}{v}{n}$ by $\stredge{m}{v\cdot\ell}{n}$ in $S$
      \EndFor
      \EndIf
      \EndIf
      \State \Return $\done\cup\{n\}$
    \EndFunction

  \end{algorithmic}
  \caption{\label{algo:outputlcp} \quad
    Compute and output the longest common prefix of words in $S$,
    and remove it from all branches of $S$.}
\end{algorithm}

Let $\maxoutdiff(u')$ be the maximal length of outputs of $T$ on $u'$
to which the longest common prefix has been removed: $ \maxoutdiff(u') =
\max_{v\in\reach(u')} |v| - |\lcpin {u'} T| $.
We prove the following complexity result:

\begin{prop}\label{prop:complexity-eval} Given an \fvpt $T$,
  one can build in \ptime  a Turing transducer, denoted
  $\evalTT(T)$, which computes $\inter{T}$, and such that,
  after reading a prefix $u'$ of a well-nested word $u\in \Sigma^*$,
  uses space in $O((\hc(u')+1)\cdot\maxoutdiff(u'))$ on the working tape.
\end{prop}

\begin{proof}
  The first step of Algorithm \eval is the reduction of the \fvpt,
  in polynomial time~\cite{Caralp201513}. This procedure
  eliminates runs starting in the initial configuration but which can not
  be completed into accepting runs. As a consequence, the value of
  $\maxoutdiff(u')$ can only decrease.
  Given a reduced \fvpt $T$,
  Algorithm \eval uses, after reading a prefix $u'$ of an input word $u$,
  space in $O(|Q|^2\cdot|\Gamma|^2\cdot (\hc(u')+1)\cdot\maxoutdiff(u'))$
  on the working tape.
  Indeed, the depth of the DAG obtained after reading $u'$ is
  $\hc(u')+1$,
  each level has at most $|Q|\cdot|\Gamma|$ nodes,
  and each edge is labelled with a word of length less than
  $\maxoutdiff(u')$ (as each edge participates in a useful \dconf,
  $T$ being reduced).
%
\end{proof}


\section{Height Bounded Memory Evaluation}

As we have seen,
bounded memory is too restrictive
in the context of nested words as it does not allow
one to process well-nested words of unbounded height.
In this section, we define
a notion of bounded memory which takes into account
the height of the input word.

\subsection{HBM transductions}

\begin{defi}\label{def:hbm}
A (functional) transduction $f : \Sigma^*\rightarrow \Sigma^*$ is
\emph{height bounded memory (HBM)} if there exists a function
$\lambda:\Nat\rightarrow \Nat$ such that $f$ is computable
by some Turing transducer $M$ that runs in space complexity
at most $\lambda(\h(u))$.
\end{defi}

\newcontent{Note that this definition ensures that the Turing transducer cannot store
the whole input word on the working tape in general, because the
length of an input word is not necessarily bounded by some function of
its height.}

\begin{exa}
\newcontent{
Given some alphabet $A$, one can encode any $A$-labelled ranked tree
into some nested word over the set of internal
symbols $A$, the set of call symbols $\{c_a\mid a\in A\}$ and the set
of return symbols $\{r_a\mid a\in A\}$ naturally by considering a
depth-first traversal of the tree. For instance, $f(g(a,b),a)$ is
encoded as $c_{f}c_{g}abr_{g}ar_{f}$. Any functional \vpt $T$ whose domain is
included in the set of encodings of $A$-labelled ranked trees of arity
at most $k$, where $k$ is a fixed constant, is in HBM\@.
Indeed, the length of any input word $u$ is then at most
exponential in $\h(u)$, and the number of runs of $T$
on $u$ is at most exponential in $|u|$, hence the result.}

Another example of \vpt-transduction is that of
\Figure~\ref{fig:eval-compact}~(a): it is not in BM, but is in HBM\@: the
stack content suffices (and is necessary) to determine the output.
\end{exa}

We have seen that a functional transduction defined by an \fst $T$ is
BM iff $T$ is sequentializable. We give an example illustrating
that for \vpts, being sequentializable is too strong to
characterize HBM\@. Consider the \vpt of \Figure~\ref{fig:vpt3}
defined by the plain
arrows. The transduction it defines is in HBM
as its domain only contains ranked trees.
However it is not sequentializable,
as the transformation of $c$ into $a$ or $b$ depends on the last
return.

When the structured alphabet contains only internal letters, HBM and
BM coincide, thus it is undecidable whether a pushdown transducer is
HBM\@.
In the remainder of this section, we prove that HBM is
decidable for \fvpts.

\subsection{Horizontal twinning property}

As we have seen in Section~\ref{sec:pb}, BM functional \fst-transductions
are characterized by the twinning property.
We introduce a similar
characterization of HBM \fvpts-transductions, called the
\emph{horizontal twinning property} (HTP).
 Intuitively, the HTP requires that two runs on the same input cannot
 accumulate increasing output delay on loops on well-nested input words.

\begin{defi}
  Let $T$ be an \fvpt. $T$ satisfies the \emph{horizontal
    twinning property} (HTP) if for all $u_1,u_2,v_1,v_2,w_1,w_2\in\Sigma^*$
  such that $u_2$ is well-nested,
  for all $q_0,q'_0\in I$, for all $q,q'\in Q$,
  and for all $\sigma,\sigma'\in \Gamma^*$ such that $(q,\sigma)$
  and $(q',\sigma')$ are co-accessible,

\noindent if $
\left\{\begin{array}{llllllll}
  (q_0,\bot) & \xrightarrow{u_1/v_1} & (q,\sigma) &  \xrightarrow{u_2/v_2} &
  (q,\sigma) \\

  (q'_0,\bot) & \xrightarrow{u_1/w_1} & (q',\sigma') &  \xrightarrow{u_2/w_2} &
  (q',\sigma') \\
\end{array}\right.
\quad$ then $\del(v_1,w_1) = \del(v_1v_2,w_1w_2)$.
\end{defi}

\begin{exa}\label{exple:vpt2}
  Consider the VPT of \Figure~\ref{fig:vpt3} (including dashed
  arrows). It does not satisfy the HTP, as the delays increase when
  looping on $crcr..$. Without the dashed transitions, the HTP is
  trivially satisfied. Indeed, for any input word there is no loop
  between configurations, that is
  any two reached configurations differ either on the stack or on the state.

\end{exa}

\subsection{Deciding HTP}

\begin{lem}\label{lm:notHTP}
Let $T$ be an \fvpt. $T$ does not satisfy the HTP
iff there exist two runs satisfying the premises of the HTP
such that either~$(i)$ $|v_2|\neq |w_2|$ or~$(ii)$ \newcontent{$v_2w_2\neq \epsilon$} and there exists $1\leq i \leq \min(|v_1|,|w_1|)$ 
such that $v_1[i]\neq w_1[i]$. 
\end{lem}

\begin{proof}
\newcontent{
First, we prove the 'if' direction. Let us take states, words and stack contents as in the premises
of the HTP:}
\begin{equation}\label{eq:one}
\left\{\begin{array}{llllllll}
  (q_0,\bot) & \xrightarrow{u_1/v_1} & (q,\sigma) &  \xrightarrow{u_2/v_2} &
  (q,\sigma) \\

  (q'_0,\bot) & \xrightarrow{u_1/w_1} & (q',\sigma') &  \xrightarrow{u_2/w_2} &
  (q',\sigma') \\
\end{array}\right.\tag{1} 
\end{equation}
\newcontent{
and suppose that $(i)$ or $(ii$) hold. By iterating the loop, i.e., by
considering input $u_2^i$ for all $i\geq 1$, we can rewrite the latter
pattern as:}
\begin{equation}\label{eq:star}
\left\{\begin{array}{llllllll}
  (q_0,\bot) & \xrightarrow{u_1u_2^{i-1}/v_1v_2^{i-1}} & (q,\sigma) &  \xrightarrow{u_2/v_2} &
  (q,\sigma) \\

  (q'_0,\bot) & \xrightarrow{u_1u_2^{i-1}/w_1w_2^{i-1}} & (q',\sigma') &  \xrightarrow{u_2/w_2} &
  (q',\sigma') \\
\end{array}\right. \tag{$\star$} 
\end{equation}
\newcontent{
If $(i)$ holds, then the difference between the lengths of $v_1v_2^{i-1}$
and $w_1w_2^{i-1}$ gets arbitrarily large when $i$ increases, and
cannot be compensated by just outputting one more $v_2$ and $w_2$. Hence,
there must necessarily exist $i$ such that
$\del(v_1v_2^{i-1},w_1w_2^{i-1})\neq
\del(v_1v_2^{i},w_1w_2^i)$. The runs~\eqref{eq:star} witness the
non-satisfiability of the HTP\@. Similarly, if $(ii)$ holds, since
$v_2w_2\neq \epsilon$, at least one of $v_2,w_2$ is non-empty, hence
by iterating the loop the delays will accumulate after the mismatch
between $u_1$ and $v_1$. In other words, we have
$\del(v_1v_2^{i-1},w_1w_2^{i-1})\neq \del(v_1v_2^{i},w_1w_2^i)$ for
all $i\geq 1$, hence again witnessing the non-satisfiability of the
HTP\@. \newline
Conversely, suppose that the HTP does not hold for the runs depicted
in~\eqref{eq:one}, i.e. $\del(v_1,w_1)\neq \del(v_1v_2,w_1w_2)$. Assume that
$|v_2|=|w_2|$. Since $\del(v_1,w_1)\neq \del(v_1v_2,w_1w_2)$, we
necessarily have $v_2w_2\neq \epsilon$ and since $|v_2|=|w_2|$ we get
that both $v_2$ and $w_2$ are non-empty. Suppose that there exists $i$
such that there is a mismatch between $v_1v_2^i$ and $w_1w_2^i$, then
we are done, by taking as witness of the right statement of the lemma
the following runs:}
    \[
    \left\{\begin{array}{llllllll}
             (q_0,\bot) & \xrightarrow{u_1u_2^i/v_1v_2^i} & (q,\sigma) &  \xrightarrow{u_2/v_2} &
                                                                                        (q,\sigma) \\

             (q'_0,\bot) & \xrightarrow{u_1u_2^i/w_1w_2^i} & (q',\sigma') &  \xrightarrow{u_2/w_2} &
                                                                                           (q',\sigma') \\
           \end{array}\right.
       \]
\newcontent{Otherwise, for all $i$, there is no mismatch between $v_1v_2^i$ and
$w_1w_2^i$. This implies that by iterating $\omega$-times the loop we
obtain the equality $v_1v_2^\omega = w_1w_2^\omega$. Wlog assume that
$v_1$ is a prefix of $w_1$, i.e. $w_1 = v_1w'_1$ for some $w'_1$ (the other
case is symmetric). Then we get $v_2^\omega = w'_1w_2^\omega$. Using
simple arguments of word combinatorics (based on Fine and Wilf's
theorem, which can be applied since $v_2$ and $w_2$ are non-empty, see for instance Lemma 3
of~\cite{DBLP:journals/corr/abs-1002-1443}), we
get that there exist two words $t_1,t_2$ and $\alpha>0$, $\beta\geq
0$, such that $v_2 = {(t_1t_2)}^\alpha$, $w_2 = {(t_2t_1)}^\alpha$ and
$w'_1 = {(t_1t_2)}^\beta t_1$. Therefore, $\del(v_1,w_1) =
\del(v_1,v_1w'_1) = (\epsilon, w'_1) = (\epsilon, {(t_1t_2)}^\beta
t_1)$. On the other hand, we have
\begin{align*}
\del(v_1v_2,w_1w_2)
&= \del(v_2,w'_1w_2) = \del({(t_1t_2)}^\alpha,{(t_1t_2)}^\beta t_1{(t_2t_1)}^\alpha) \\
&= \del({(t_1t_2)}^\alpha,{(t_1t_2)}^{\beta+\alpha} t_1) = (\epsilon, {(t_1t_2)}^\beta t_1) = \del(v_1,w_1),
\end{align*}
contradicting our assumption.}
\end{proof}

\begin{prop}\label{prop:HTP}
  The HTP is decidable in \conptime for \fvpts.
\end{prop}

\begin{proof}
\newcontent{
  Let $T$ be an \fvpt. We reduce HTP decidability to checking the emptiness
  of a non-deterministic reversal-bounded pushdown counter machine $M$, in polynomial
  time. Such a machine is a pushdown automaton extended with counters
  which can be incremented, decremented, and tested to zero. Counters
  can be in two modes, either increasing or decreasing. A
  \emph{reversal} is a change of mode. Given a fixed constant $r$ and
  a fixed number of counters $k$, the emptiness problem of
  any pushdown $k$-counter machine
  whose runs (on any input) make at most $r$ reversals, is decidable
  in \conptime~\cite{vpts10}. \newline
  In our reduction, one only needs to take
  $r=1$ and $k=2$. Our pushdown counter machine $M$
  accepts any word of the form
  $u=u_1\#u_2\#u_3\#u_4$, where $\#$ is a special separator symbol,
  such that there exist runs of $T$ of the form:}
  \[
  \left\{\begin{array}{llllllll}
           (q_0,\bot) & \xrightarrow{u_1/v_1} & (q,\sigma) &  \xrightarrow{u_2/v_2} &
                                                                                      (q,\sigma)
           & \xrightarrow{u_3/v_3} (q_f,\bot)\\

           (q'_0,\bot) & \xrightarrow{u_1/w_1} & (q',\sigma') &  \xrightarrow{u_2/w_2} &
                                                                                         (q',\sigma')
           & \xrightarrow{u'_3/w_3} (q'_f,\bot)\\
         \end{array}\right.
     \]
\newcontent{     where $q_f,q'_f$ are accepting, and such that
  either $(i)$ $|v_2|\neq |w_2|$ or $(ii)$ $v_2w_2\neq\epsilon$ and there exists $1\leq i \leq \min(|v_1|,|w_1|)$
such that $v_1[i]\neq w_1[i]$ (\ie, the HTP does not hold by
Lemma~\ref{lm:notHTP}). The words $u_3$ and $u'_3$ are used to check
that the configurations $(q,\sigma)$ and $(q',\sigma')$ are
co-accessible. Therefore the HTP holds if and only if no word is accepted by the
automaton $M$.  \newline
When reading $u$, $M$ non-deterministically guesses whether condition
$(i)$ or condition $(ii)$ holds. For each of them, it will simulate
the behaviour of $T$ (ignoring the outputs), by guessing
non-deterministically a run of $T$ on $u_1u_2u_3$ and a run on $u_1u_2u'_3$
(hence the states of $M$ contain pairs of states of $T$), making sure
that those runs loop on $u_2$. To do so, when reading the first $\#$,
the states $(q,q')$ reached by $T$ on the two runs are remembered, then $u_2$
is checked to be well-nested (using the pushdown stack), and once the second $\#$ is met, it
suffices for $M$ to verify that the pairs of states reached so far is
$(q,q')$, otherwise the run of $M$ rejects. \newline
Let us now explain how condition $(i)$ is checked. It suffices to have
two counters $c_1,c_2$ counting the length of $v_2$ and $w_2$
respectively. To do so, the two counters $c_1,c_2$ are initially set
to $0$ and after reading the first $\#$, when guessing the two runs of
$T$ on $u_2$, the lengths of the outputs of any two respective simulated
transitions of $T$ are added to $c_1$ and $c_2$. When the second $\#$
is met, $M$ checks whether $c_1\neq c_2$, using $\epsilon$ transitions (whose use is allowed
in the model) to decrement in parallel $c_1,c_2$ until the point where
one of them reaches $0$. At this point, it suffices to check that the
other one is not zero (in which case the two lengths are different),
otherwise $M$ rejects. \newline
We now detail how to check condition $(ii)$. Initially, $c_1$ and
$c_2$ are filled with an arbitrary value $i$. This can be done again by
using some $\epsilon$-loop which increments both
counters in parallel. Then, in parallel to simulating two runs of $T$,
$M$ decrements $c_1,c_2$ by the length of the outputs of the
transitions taken by the two simulated runs respectively. When one of
them reaches $0$, say $c_1$, it means that that the output $v_1$ of
the first run has reached position $i$, it suffices to store the
current symbol $v_1[i]$ in the state of $M$. When later on the other counter, say $c_2$, reaches $0$,
it means that the second run of $T$ has reached position $i$ of $v_2$,
and $M$ can therefore check whether $v_1[i]\neq v_2[i]$. }
\end{proof}

\subsection{Deciding HBM}

We now show that HTP characterizes HBM \fvpts-transductions.

\begin{thm}\label{thm:vptsboundedmem}\label{thm:bm}
  Let $T$ be an \fvpt. $\inter{T}$ is HBM iff the HTP holds for
  $T$, which is decidable in \conptime. In this case, the Turing
  transducer $\evalTT(T)$ runs, on an input stream $u$, in space
  complexity exponential in the height of $u$.
\end{thm}

We can state more precisely the space complexity of
$\evalTT(T)$ when $T$ is reduced. In this case, it is in
$O(3{(\h(u)+1)}^3\cdot |Q|^{2(\h(u)+1)} \cdot M)$,
where $M=\max \{|v| \hmid (q,a,v,\gamma,q')\in\delta\}$.


\begin{proof}
\newcontent{
  Let $T$ be an \fvpt.
  If $\inter{T}$ is HBM, then the HTP holds for $T$ by Lemma~\ref{lem:hbm-implies-htp} (proved in this section).
  Conversely, if $T$ satisfies the HTP,
  thanks to Theorem~6.1 of~\cite{Caralp201513},
  we build an equivalent reduced \fvpt $T'$
  in polynomial time (more precisely, we use the construction \textsf{reduce}
  of~\cite{Caralp201513}).
  In this construction, the states and the stack symbols of $T'$
  are obtained from those of $T$ by enriching them with a state of $T$.
  In addition, given a run in $T'$, one recovers a run in $T$ (with the
  same input and output words) when projecting
  away this additional component. As a consequence, the fact that $T$
  satisfies the HTP implies that $T'$ also does.
  We thus assume now that $T$ is reduced.
}

Then we apply Lemma~\ref{lm:HTP-delay} (proved in this section) which
bounds the maximal difference between outputs of $T$ on a prefix $u'$
of the input $u$:
$\maxoutdiff(u') \leq 3{(\h(u')+1)}^2|Q|^{2(\h(u')+1)}M$.
Proposition~\ref{prop:complexity-eval} gives the complexity of the
evaluation algorithm:
$O((\hc(u')+1)\cdot\maxoutdiff(u'))$ on the working tape
after reading a prefix $u'$ of $u$.
We know that $\hc(u') \le \h(u')\le\h(u)$, so the space is in
$O(3{(\h(u)+1)}^3\cdot |Q|^{2(\h(u)+1)} \cdot M)$,
and finally $\inter{T}$ is HBM\@.
Hence deciding HBM reduces to deciding HTP, and this is in \conptime by
Proposition~\ref{prop:HTP}.
\end{proof}

\begin{lem}\label{lem:hbm-implies-htp}
  Let $T$ be an \fvpt. If $\inter{T}$ is HBM, then the HTP holds
  for $T$.
\end{lem}

\begin{proof}
  \ignore{ 
  By definition of HBM and BM, if $\inter{T}$ is HBM and there
  exists $K\in\Nat$ such that for all $u\in\dom(T)$, $h(u)\leq
  K$, then $\inter{T}$ is BM\@.
  }

  Suppose that the HTP does not hold for $T$. Therefore there are words
  \[
    u_1,u_2,u_3,u'_3,v_1,v_2,v_3,w_1,w_2,w_3,w_3\in\Sigma^*,
\]
  stacks $\sigma,\sigma'\in\Gamma^*$ and
  states $q,q'\in Q$, $q_0,q'_0\in I$ and
  $q_f,q'_f\in F$ such that:
  \[
  \left\{\begin{array}{llllllll}
  (q_0,\bot) & \xrightarrow{u_1/v_1} & (q,\sigma) &  \xrightarrow{u_2/v_2} &
  (q,\sigma) & \xrightarrow{u_3/v_3} & (q_f,\bot)\\

  (q'_0,\bot) & \xrightarrow{u_1/w_1} & (q',\sigma') &  \xrightarrow{u_2/w_2} &
  (q',\sigma') & \xrightarrow{u'_3/w_3} & (q'_f,\bot)\\
  \end{array}\right.
  \]
  and $\del(v_1,w_1) \neq \del(v_1v_2,w_1w_2)$. Let $K =
  \text{max}(\h(u_1u_2u_3),\h(u_1u_2u'_3))$. By definition of
  $\fst(T,K)$ (states are configurations of $T$) and Definition~\ref{def:tp2},
  the twinning property for \fsts does not hold for $\fst(T,K)$.
  Therefore, by~\cite{DBLP:journals/tcs/Choffrut77} (see Lemma~\ref{prop:bmfst}),  $\fst(T,K)$ is not sequentializable.
  By Proposition~\ref{prop:bmfst}, $\inter{\fst(T,K)}$ is not BM\@.
  Therefore $\inter{T}$
  is not HBM, otherwise $\inter{T}$ could be evaluated in space
  complexity $f(\h(u))$ on any input word $u$, for some function
  $f$. This corresponds to bounded memory if we fix the height of the
  words to $K$ at most.
\end{proof}

For the converse, we can apply the evaluation algorithm of Section~\ref{sec:algo}, whose complexity depends on the maximal delay between
all the candidate outputs of the input word.
This maximal delay is exponentially bounded by the
height of the word.

\olivier{J'avais ajout\'e ce lemme pour corriger la preuve du Lemme~\ref{lm:HTP-delay}}
In order to prove this result, we introduce a notion of arity
by analogy with trees that can be encoded by well-nested words.
The \emph{arity} $\ar$ of a well-nested word is inductively defined by:
$\ar(i) = 1$ if $i\in\Sigma_\iota$,
$\ar(uv) = \ar(u) + \ar(v)$ if $u$ and $v$ are well-matched,
and $\ar(cur) = 1$ if $u$ is well-matched,
and $c$ and $r$ are call and return symbols, respectively.
We say that a well-nested word $u$ is \emph{$k$-narrow}
if $\ar(v) \le k$ for all well-nested factors $v$ of $u$.

\begin{lem}\label{lem:arity-bound}
  If $u$ is a $k$-narrow well-nested word with $k\ge 2$
  then $|u|\leq 3(k^{\h(u)+1}-1)$.
\end{lem}

\begin{proof}
  We show, by induction on $\h(u)$,
  that $|u|\le \frac{2k}{k-1}(k^{\h(u)+1}-1)$.
  Note that this implies $|u|\leq 3(k^{\h(u)+1}-1)$, as $k\ge 2$.
  Let us consider the unique decomposition $u=u_1u_2\cdots u_n$
  with $\ar(u_i)=1$ for all $1\le i\le n$ and $n=\ar(u)\le k$.
  The basic case is $\h(u)=0$, \ie, every word $u_i$ is an internal symbol.
  In that case \[
    |u| = n \le k \le 2k = \frac{2k}{k-1}(k^{\h(u)+1}-1).
    \]
  In the general case, every word $u_i$ is either an internal symbol
  (and thus of length $1$), or of the form $c_{i}v_{i}r_{i}$
  where $c_i$ (resp.\ $r_i$) is a call (resp.\ return) symbol
  and $v_i$ a well-nested word such that $\h(v_i)<\h(u)$ so,
  by induction hypothesis,
  \[
    |v_i| \le \frac{2k}{k-1}(k^{\h(v_i)+1}-1) \le \frac{2k}{k-1}(k^{\h(u)}-1).
  \]
  Hence,
  \[
  |u| \le k(2 + \frac{2k}{k-1}(k^{\h(u)}-1))
  = 2k\frac{k-1+k(k^{\h(u)}-1)}{k-1}
  = \frac{2k}{k-1}(k^{\h(u)+1}-1).
  \qedhere
  \]
\end{proof}

\olivier{J'ai corrig\'e le lemme et sa preuve: \`a relire\ldots}
\begin{lem}%
\label{lm:HTP-delay}
Let $T$ be a reduced \fvpt. If the HTP holds for $T$, then for all
well-nested words $u\in\Sigma^*$, and all prefixes $u'$ of $u$, we have
\[\maxoutdiff(u') \leq 3{(\h(u')+1)}^2|Q|^{2(\h(u')+1)}M,\]
where $M=\max \{|t| \mid (q,a,t,\gamma,q')\in\delta\}$.
\end{lem}


\begin{proof}
  Let $u\in\Sigma^*$ be a well-nested word,
  and $u'\in \Sigma^*$ be a prefix of $u$.
  The proof is similar to that
  of~\cite{DBLP:journals/tcs/BealC02} for FST\@. It proceeds by
  induction on the length of $u'$.
  If $|u'|\leq 3{(\h(u')+1)}^2|Q|^{2(\h(u')+1)}$, then the result is
  trivial.
  Otherwise, consider the unique decomposition of $u'$ such that
  $u'=u_0c_1u_1c_2\cdots u_{n-1}c_{n}u_{n}$ where every $u_i$ is well-nested,
  and every $c_i$ is a call symbol. Hence $n = \hc(u') \le \h(u')$.

  Let us first consider the case where every $u_i$ is $|Q|^2$-narrow.
  If $|Q|>1$ then,
  by Lemma~\ref{lem:arity-bound},
  $|u_i| \le 3(|Q|^{2(\h(u_i)+1)}-1)$ and thus
  \[|u'| \le (n+1)(3(|Q|^{2(\h(u')+1)}-1)+1),\]
  since $\h(u_i)\le\h(u')$.
  As $n \le \h(u')$,
  \[|u'| \le 3(\h(u')+1)|Q|^{2(\h(u')+1)} \le 3{(\h(u')+1)}^2|Q|^{2(\h(u')+1)},\]
  which means that we are in the basic case.
  If $|Q|=1$ then each $u_i$ is $1$-narrow and thus of
  the form $c_1'\cdots c_\ell'\iota r_\ell' \cdots r_1'$
  with $\iota\in\{\emptyword\}\cup\Sigma_\iota$ and,
  for every $j$, $c_j'\in\Sigma_c$, $r_j'\in\Sigma_r$.
  So $|u_i| \le 2\h(u_i)+1$ and
  \[|u'| \le (n+1)(2\h(u_i)+2) \le 2{(\h(u')+1)}^2
  \le 3{(\h(u')+1)}^2|Q|^{2(\h(u')+1)}.\] So we are also in the basic case.

  Assume now that one of the $u_i$ is not $|Q|^2$-narrow,
  \ie, $\ar(u'') > |Q|^2$ for some well-nested factor $u''$ of $u_i$.
  Let $(q,\sigma,w)$, $(q',\sigma',w')\in Q\times \Gamma^*\times \Sigma^*$
  be such that there exist runs
  $\rho:  (i,\bot)\xrightarrow{u'/v}(q,\sigma)$ and
  $\rho': (i',\bot)\xrightarrow{u'/v'}(q',\sigma')$, with $i,i'\in I$,
  $v = \lcpin {u'} T\cdot w$, $v' = \lcpin {u'} T\cdot w'$,
  and such that $\maxoutdiff(u') = |w|$.
  Consider the decomposition of $u''$ into well-nested words $u_i''$
  of arity one: $u''=u_1''u_2''\cdots u_k''$,
  and consider the states $p_i$ (resp. $p_i'$) encountered in $\rho$
  (resp. $\rho'$) after reading $u_i''$.
  As $k = \ar(u'') > |Q|^2$, there exist $i,j$ such that $1\le i < j\le k$
  and $(p_i,p_i')=(p_j,p_j')$.
  Let $y=u_{i+1}''u_{i+2}''\cdots u_j''$.
  We can decompose runs $\rho$ and $\rho'$ as follows:
  \[
  \left\{\begin{array}{llllllllll}
  \rho: (i,\bot) &
  \xrightarrow{x/v_1} & (p_i,\sigma_1) & \xrightarrow{y/v_2} &
  (p_i,\sigma_1) & \xrightarrow{z/v_3} & (q,\sigma) \\

  \rho': (i',\bot) &
  \xrightarrow{x/v'_1} & (p_i',\sigma'_1) &  \xrightarrow{y/v'_2} &
  (p_i',\sigma_1') & \xrightarrow{z/v_3'} & (q',\sigma') \\
  \end{array}\right.
  \]
  In addition, we have $u'=xyz$, and $y\neq\varepsilon$ and well-nested,
  $v=\lcpin {u'} T\cdot w = v_1v_2v_3$, and
  $v'=\lcpin {u'} T\cdot w' = v_1'v_2'v_3'$.
  Moreover, $(q,\sigma)$ and $(q',\sigma')$ are co-accessible,
  as $T$ is reduced.
  By the HTP property, we obtain
  $\del(v_1v_2,v_1'v_2')=\del(v_1,v_1')$.
  By Lemma~\ref{lem:assocdelay}, this entails the equality
  \[\del(v_1v_2v_3,v_1'v_2'v'_3)
  =\del(\del(v_1v_2,v_1'v_2')\cdot (v_3,v_3'))
  =\del(\del(v_1,v_1')\cdot (v_3,v_3'))
  =\del(v_1v_3,v_1'v'_3).\]  Thus, we obtain
  \[
    \del(w,w')=\del(v,v')=\del(v_1v_2v_3,v_1'v_2'v_3')=\del(v_1v_3,v_1'v_3').
  \]
  As
  $v_1v_3$ and $v'_1v'_3$ are possible output words for the input word
  $u_1u_3$, whose length is strictly smaller than $|u'|$, we obtain
  $\maxoutdiff(u') = |w| \leq \maxoutdiff(u_1u_3)$
  and the result holds by induction.
\end{proof}

\begin{exa}[HBM is tight]
  \newcommand\encode{\ensuremath{\mathsf{encode}}\xspace}
  Theorem~\ref{thm:bm} shows that the space complexity of a \vpt in
  HBM is at most exponential. We give here an example illustrating the
  tightness of this bound.
  We describe here a transduction on trees, and use the
  well-known encoding of trees by well-nested words defined by:
  $\encode(a(t_1,\dots,t_n)) = c_a\cdot\encode(t_1)\cdots\encode(t_n)r_a$.
  The idea is to encode the tree transduction
  $\big(f(t,a) \mapsto f(t,a) \big)\cup \big(f(t,b)\mapsto f(\overline{t},b)\big)$
  by a VPT, where $t$ is a
  binary tree over $\{0,1\}$ and $\overline{t}$ is the mirror of $t$,
  obtained by replacing the $0$ by $1$ and the $1$ by $0$ in $t$.
  Thus taking the identity or the mirror depends on the second child of the
  root $f$. To evaluate this transformation in a streaming manner, one
  has to store the whole subtree $t$ in memory before deciding to
  transform it into $t$ or $\overline{t}$. The evaluation of this
  transduction cannot be done in space polynomial in the height of the input
  as there are a doubly exponential number of trees of height $n$, for
  all $n\geq 0$.
\end{exa}


\section{Online Bounded Memory Evaluation}\label{sec:twinning}

In the previous section, we have shown that a \vpt-transduction is in
HBM iff the horizontal twinning property holds. The notion of
height-bounded memory is quite permissive as for instance, any
transduction of \emph{ranked} tree linearizations (given a fixed
rank) is HBM, because in this case, the length of the tree
linearization functionally depends on the height of the tree.

In this section, we introduce a stronger constraint on
memory: the amount of memory must depend only, at each moment (i.e., at
any position in the nested word), on the current height. We call this
requirement \emph{online bounded memory} (\OBM).

\subsection{OBM transductions}

\begin{defi}
A (functional) transduction $f : \Sigma^*\rightarrow \Sigma^*$ is
\emph{online bounded memory} (\OBM) if there exists a function
$\lambda:\Nat\rightarrow \Nat$ such that $f$ is computable by a
Turing transducer $M$ satisfying the following property:
on any input word $u\in \Sigma^*$,
if $M$ has read the prefix $v$ of $u$ (but not more),
then the amount of memory of $M$ on the working tape is less than
$\lambda(\hc(v))$.
\end{defi}

\begin{exa}\label{ex:obmtrans}
Consider the transduction that maps any word of the form $c^{n}r^{n}$ to
$a^{n}c^{n}$, and any word of the form $c^{n}r'r^{n-2}r'$ to $b^{n}c^{n}$. This
transduction is \OBM\@: it suffices to compute the number $n$ of $c$ symbols
till we read a symbol $r$ or $r'$. During this phase, the memory is in
$O(\log(n))$ and depends on the current height of the word. When $r$ (resp. $r'$) is read
for the first time, the word $a^{n}c$ (resp. $b^{n}c$) is output and the
memory flushed. Then, whenever a symbol is read, nothing is stored on
memory and $c$ symbols are output whenever a symbol is read.
\end{exa}

In this section, we give an effective characterization of \OBM
transductions definable by \VPT,  using a new twinning property,
called the  \emph{matched twinning property} (or \emph{MTP}
for short). Since it is a characterization of transductions,
this property does not depend on the
\VPT that implements them: two equivalent \VPTs that
implements the same transduction both satisfy, or both do not satisfy, this twinning property.
Another appealing property of \OBM, compared to \HBM, is
that the maximal amount memory needed when running the algorithm of
Section~\ref{sec:algo} is at most \emph{quadratic} in the current height of
the input nested word, while it is \emph{exponential} for \HBM transductions,
and this latter bound is tight. In other words, any \OBM transduction
defined by a \vpt can be evaluated with quadratic memory in the height
of the input nested word.

\subsection{Matched twinning property (MTP)}

\begin{figure}[t]
\centering

\begin{tikzpicture}[scale=.1]

  \draw (0,0)
  to [out=0,in=180] (7,5)
  to [out=0,in=180] (10,3)
  to [out=35,in=215] (15,11) node(pq1){$\bullet$} node[above left]{$(p,p')$}
  to [out=0,in=180] (20,18)
  to [out=330,in=250] (30,32) node(pq2){$\bullet$} node[above left]{$(p,p')$}
  to [out=70,in=180] (45,34)
  to [out=0,in=120] (60,32) node(ppqp1){$\bullet$} node[above right]{$(q,q')$}
  to [out=290,in=120] (75,11) node(ppqp2){$\bullet$} node[above right]{$(q,q')$};

  \draw[->] (0,0) -- (0,40);
  \draw[->] (0,0) -- (95,0);
  \draw (0,42) node{height};
  \draw (100,0) node{input};

  \draw[dashed] (0,32) -- (60,32);
  \draw[dashed] (0,11) -- (75,11);
  \draw[dotted] (15,0) -- (15,11);
  \draw[dotted] (30,0) -- (30,32);
  \draw[dotted] (60,0) -- (60,32);
  \draw[dotted] (75,0) -- (75,11);

  \draw ( 7,-3) node{$u_1$};
  \draw (23,-3) node{$u_2$};
  \draw (45,-3) node{$u_3$};
  \draw (68,-3) node{$u_4$};

\end{tikzpicture}

\caption{Premisses of the matched twinning property (\MTP)}\label{fig:mtp}
\end{figure}
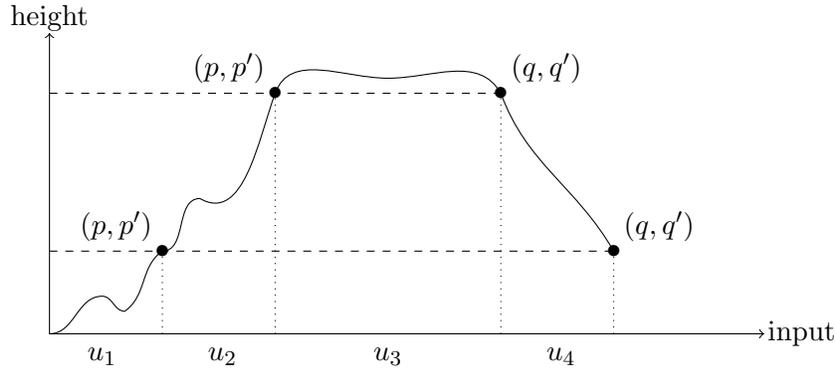

The matched twinning property is a strenghtening of
the horizontal twinning property obtained by adding some new delay
constraints on the well-matched loops. Intuitively, the
\MTP requires that two runs on the same input cannot accumulate
increasing output delay on well-matched loops. They can accumulate
delay on loops with increasing stack but this delay has to be caught
up on the matching loops with descending stack.  We show that this property is
decidable, and that sequential \vpts satisfy it.
Therefore the class of \OBM \VPT-transductions \emph{subsumes} the class of
sequentializable \VPTs.
%
%
We illustrate the following definition in Figure~\ref{fig:mtp}.


\begin{defi} 
  Let $T = (Q,I,F,\Gamma, \delta)$ be an \fvpt.
  $T$ satisfies the \emph{matched twinning property} (\MTP) if for all
  $u_i,v_i,w_i\in\Sigma^*$ ($i\in\{1,\dots,4\}$) such that $u_3$ is
  well-nested, and $u_2u_4$ is well-nested, for all $i,i'\in I$,
  for all $p,q,p',q'\in Q$, and for all $\sigma_1,\sigma_2\in
  \bot.\Gamma^*$, for all $\sigma'_1,\sigma_2'\in \Gamma^*$,
 such that
  $(q,\sigma_1)$ and $(q',\sigma_2)$ are co-accessible:

\smallskip
\noindent if $\quad
\left\{\begin{array}{llllllllllllllllll}
  (i,\bot) & \xrightarrow{u_1/v_1} & (p,\sigma_1) &  \xrightarrow{u_2/v_2} &
  (p,\sigma_1\sigma_1') &  \xrightarrow{u_3/v_3} & (q,\sigma_1\sigma'_1) &
\xrightarrow{u_4/v_4} & (q,\sigma_1) \\

  (i',\bot) & \xrightarrow{u_1/w_1} & (p',\sigma_2) &  \xrightarrow{u_2/w_2} &
  (p',\sigma_2\sigma'_2) &  \xrightarrow{u_3/w_3} & (q',\sigma_2\sigma'_2) &
\xrightarrow{u_4/w_4} & (q',\sigma_2) \\
\end{array}\right.$

\smallskip
\noindent then $\del(v_1v_3,w_1w_3) =
\del(v_1v_2v_3v_4,w_1w_2w_3w_4)$. We say that a \vpt $T$ is
\emph{twinned} whenever it satisfies the \MTP.
\end{defi}

Note that any twinned \vpt also satisfies the HTP (with $u_3 = u_4 =
\epsilon$).

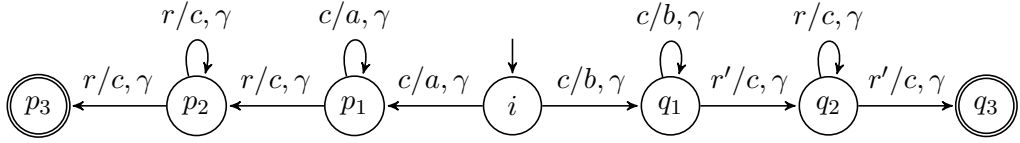
\begin{figure}[t]
\centering
\begin{tikzpicture}[->,>=stealth',shorten >=1pt,auto,node distance=2.1cm,
                    semithick]

\node[state,accepting] (p3) {$p_3$};
\node[state]           (p2) [right of=p3] {$p_2$};
\node[state]           (p1) [right of=p2] {$p_1$};
\node[state,initial above]   (i)  [right of=p1] {$i$};
\node[state]           (q1) [right of=i]  {$q_1$};
\node[state]           (q2) [right of=q1] {$q_2$};
\node[state,accepting] (q3) [right of=q2] {$q_3$};


\path (i)  edge [above]      node {$c/a, \gamma$} (p1)
      (p1) edge [loop above] node {$c/a, \gamma$} (p1)
           edge [above]      node {$r/c, \gamma$} (p2)
      (p2) edge [loop above] node {$r/c, \gamma$} (p2)
           edge [above]      node {$r/c, \gamma$} (p3);

\path (i)  edge [above]      node {$c/b, \gamma$} (q1)
      (q1) edge [loop above] node {$c/b, \gamma$} (q1)
           edge [above]      node {$r'/c, \gamma$} (q2)
      (q2) edge [loop above] node {$r/c, \gamma$} (q2)
           edge [above]      node {$r'/c, \gamma$} (q3);


\end{tikzpicture}
\caption{A non-sequentializable \vpt that defines an \OBM-transduction.}\label{fig:vpt4}
\end{figure}


\begin{exa}[MTP does not imply sequentializable]\label{exple:vpt3}
  The \vpt of \Figure~\ref{fig:vpt3} with plain arrows does not satisfy
  the \MTP, as the delay between the two branches increases when
  iterating the loops. Consider now the \vpt of
  \Figure~\ref{fig:vpt4}. It implements the transduction of Example~\ref{ex:obmtrans}.
  It is obviously twinned, as we cannot
  construct two runs on the same input which have the form given in
  the premises of the \MTP. However this transducer is not
  sequentializable, as the output on the call symbols cannot be
  delayed to the matching return symbols.
\end{exa}

\subsection{Deciding MTP} \newcontent{As for the HTP, one can decide the \MTP using a reduction
to the emptiness of a reversal-bounded pushdown counter
machines. First, one needs a technical lemma about the delays of
iterated words, whose proof strongly relies on some word combinatorics
result by Saarela~\cite{Saarela15}.}

\begin{lem}\label{lem:diffcont}
  Let $v_1,\ldots,v_{2n+1}$ and  $w_1,\ldots,w_{2m+1}$ be two sequences of finite words over $\Sigma$,
  for two positive integers $n$ and $m$. We define, for all $i \geq 0$, the two following words: 
  \[
  \begin{array}{ll}
  V_i  & =  v_1{(v_2)}^i v_3 {(v_4)}^i \ldots v_{2n-1}{(v_{2n})}^i v_{2n+1}\\
  W_i & =  w_1{(w_2)}^i w_3 {(w_4)}^i \ldots w_{2m-1}{(w_{2m})}^i w_{2m+1}
  \end{array}
  \]
  If $\del(V_0,W_0)\neq \del(V_1,W_1)$, then for any $i\geq 0$, the set
  $\{j\geq 0 \mid \del(V_i,W_i) = \del(V_j,W_j)\}$ is finite.
\end{lem}

\begin{proof}
The proof relies on a recent result of word combinatorics proved by
Saarela in~\cite{Saarela15}. It states that if the equality
\[
\alpha_0 {(\alpha_1)}^i \alpha_2 \dots
\alpha_{2n-1}{(\alpha_{2n})}^i\alpha_{2n+1} =
\beta_0 {(\beta_1)}^i \beta_2 \dots
\beta_{2m-1}{(\beta_{2m})}^i\beta_{2m+1}
\]
holds for $m+n$ values of $i$, then it holds for all $i\geq 0$, where
$n,m\geq 0$ and the $\alpha_j,\beta_j$ are  arbitrary words.
To simplify notations and case study, we work in the free group,
i.e., the free monoid extended with the inverse $u^{-1}$, quotiented by
the equalities $\sigma^{-1}\sigma = \sigma\sigma^{-1}=\epsilon$ for
all $\sigma$.

First observe that w.l.o.g., we can assume that $v_{2n}$ and $w_{2m}$
are non-empty. Indeed, if one of them is empty, say $v_{2n} =
\epsilon$, we set $v_{2n-1}$ to $v_{2n-1}v_{2n+1}$.

We proceed by contradiction and assume that there exists
a delay $d$ such that $d=\del(V_i,W_i)$ for infinitely many $i$'s. We fix such a $d$ and we let $J\subseteq \mathbb{N}$
be the set of indices $i$ such that $d=\del(V_i,W_i)$. We write
$d=(x,y)$ and distinguish two cases:
\begin{enumerate}
\item if $x=\epsilon$ or $y=\epsilon$, for simplicity suppose that $x=\epsilon$, the other case being symmetric.
  For all $i\in J$, we have $W_i = V_i y$. Since $J$ is infinite, we
  can use Saarela's result to conclude that $W_i = V_i y$ for all
  $i\geq 0$. In particular, we get $W_0 = V_0 y$ and $W_1 =
  V_1y$. Hence, $\del(W_0,V_0) = (\epsilon,y) = \del(W_1,V_1)$ which
  is a contradiction.

\item if $x\neq\epsilon$ and $y\neq\epsilon$, we again distinguish three
  cases:
  \begin{enumerate}
    \item if $|x|\leq |v_{2n+1}|$, then $v_{2n+1} = zx$ for some
      $z$. By definition of $x$ and $y$, we know that for all $i\in
      J$, there exists $\ell_i\in\Sigma^*$ such that
      \[
      V_i = \ell_i x = v_1{(v_2)}^i v_3 {(v_4)}^i \ldots v_{2n-1}{(v_{2n})}^i
      z x\qquad\text{and}\qquad W_i = \ell_i y
      \]

      Hence, we obtain:
      \[
\begin{array}{llllllll}
      & V_i x^{-1} y & = & v_1{(v_2)}^i v_3 {(v_4)}^i \ldots v_{2n-1}{(v_{2n})}^i
  z y  & \\ & & = & W_i \\ & & = & w_1{(w_2)}^i w_3 {(w_4)}^i \ldots w_{2m-1}{(w_{2m})}^i
  w_{2m+1}
\end{array}
      \]
      for all $i\in J$. By Saarela's result and since $J$ is infinite,
      the above equation holds for all $i\geq 0$. Now, for all $i\geq
      0$, let $(x_i, y_i)
      = \Delta(V_i,W_i)$ and $\ell_i$ such that $V_i = \ell_i x_i$ and
      $W_i = \ell_i y_i$. Then, for all $i\geq 0$, $y_i = \ell_i^{-1}
      W_i = \ell_i^{-1}V_i x^{-1}y = \ell_i^{-1} \ell_i x_i x^{-1} y =
      x_i x^{-1} y$, hence $x_i^{-1} y_i = x^{-1}y$ for all $i\geq
      0$. By definition of the delay, $x$ and $y$ (which are
      both non-empty by assumption), start with different
      symbols. Hence $x^{-1}y \in
      {(\Sigma^{-1})}^{|x|}\Sigma^{|y|}$. Since $|x|,|y|>0$, it cannot
      be the case that $x_i$ or $y_i$ is empty. Hence, they both start
      with different symbols (as delays), and we get $x_i = x$ and $y_i = y$ for
      all $i\geq 0$. In particular, it means that $\del(V_0,W_0) =
      \del(V_1,W_1)$.

    \item if $|y|\leq|w_{2m+1}|$. This case is symmetric to the case $|x|\leq |v_{2n+1}|$.

    \item if $|x|> |v_{2n+1}|$ and $|y|>|w_{2m+1}|$. Since
      $\Delta(V_i,W_i) = (x,y)$ for infinitely many $i$, by taking $i\in J$ sufficiently
      large, we have for some $k,k'$ which can be assumed to be greater
      than $\text{max}(|v_{2n}|,|w_{2m}|)$ (the reason why we take
      such values will be clear later), and some $v',w'\in\Sigma^*$
      such that $v'$ and $w'$ are prefixes of $v_{2n}$ and $w_{2m}$
      respectively:
      \[
      \qquad\qquad\qquad
      \begin{array}{rcl}
      V_i & = & v_1{(v_2)}^i v_3 {(v_4)}^i \ldots v_{2n-1}{(v_{2n})}^{k}v' x
        \\
      W_i & = & w_1{(w_2)}^i w_3 {(w_4)}^i \ldots w_{2m-1}{(w_{2m})}^{k'}w' y
        \\
        v_1{(v_2)}^i v_3 {(v_4)}^i \ldots v_{2n-1}{(v_{2n})}^{k}v' & = &
                                                                 w_1{(w_2)}^i w_3 {(w_4)}^i \ldots w_{2m-1}{(w_{2m})}^{k'}w' \\
      \end{array}
      \]
      In other words, the first letter of $x$ and $y$ corresponds to
      the first mismatch between $V_i$ and $W_i$. Moreover,
      $v'x = v_{2n}^j x'$ for some $j,x'$ and $w'y = w_{2m}^{j'} y'$ for
      some $j',y'$ and since $|x|> |v_{2n+1}|$ and $|y|>|w_{2m+1}|$, we
      necessarily have $j,j'>0$. Since $v'$ and $w'$ are prefixes of
      $v_{2n}$ and $w_{2m}$ respectively, we get the existence of two
      different symbols $a,b\in\Sigma$ and two words $v'', w''$ such
      that $v_{2n} = v'av''$ and $w_{2m} =w'bw''$. Replacing these
      values in the third equation above, we get:
        \[v_1{(v_2)}^i v_3 {(v_4)}^i \ldots v_{2n-1}{(v'av'')}^{k}v' =
                                                                 w_1{(w_2)}^i w_3 {(w_4)}^i \ldots w_{2m-1}{(w'bw'')}^{k'}w' \]
                                                                 which can be rewritten into
        \[v_1{(v_2)}^i v_3 {(v_4)}^i \ldots v_{2n-1}v'{(av''v')}^{k} =
                                                                 w_1{(w_2)}^i w_3 {(w_4)}^i \ldots w_{2m-1}w'{(bw''w')}^{k'}\]

    Let $U = v_1{(v_2)}^i v_3 {(v_4)}^i \ldots v_{2n-1}v'{(av''v')}^{k}$ and
    $\ell$ its length.
    We have $U[\ell-j|v_{2n}|+1] = a$ for all $1\leq j\leq
    k$. Similarly, we have $U[\ell-j'|w_{2m}|+1] = b$ for all $1\leq
    j'\leq k'$. By taking $j = |w_{2m}|$ and $j' = |v_{2n}|$ (which is
    possible since $k,k'\geq \text{max}(|v_{2n}|,|w_{2m}|)$), we get a
    contradiction, as $a\neq b$.
    \qedhere
  \end{enumerate}
\end{enumerate}
\end{proof}

\newcontent{
\noindent
The following lemma gives an alternative characterisation of the \MTP which
can be exploited to decide the \MTP using counter pushdown machines.}

\begin{lem}\label{lem:mtpcharac}
  Let $T = (Q,I,F,\Gamma, \delta)$ be an \fvpt. $T$ does not satisfy
  the \MTP iff there exist words $u_i,v_i,w_i$, $i=1,\dots,4$,
  states $i,i'\in I$, $p,q,p',q'\in Q$, stack contents
  $\sigma_1,\sigma_2\in \bot.\Gamma^*$ and $\sigma'_1,\sigma'_2\in
  \Gamma^*$ such that the following conditions hold:
  \begin{enumerate}
    \item $(q,\sigma_1)$ and $(q',\sigma_2)$ are co-accessible, $u_3$
      and $u_2u_4$ are well-nested,

\item $\left\{\begin{array}{llllllllllllllllll}
  (i,\bot) & \xrightarrow{u_1/v_1} & (p,\sigma_1) &  \xrightarrow{u_2/v_2} &
  (p,\sigma_1\sigma_1') &  \xrightarrow{u_3/v_3} & (q,\sigma_1\sigma'_1) &
\xrightarrow{u_4/v_4} & (q,\sigma_1) \\

  (i',\bot) & \xrightarrow{u_1/w_1} & (p',\sigma_2) &  \xrightarrow{u_2/w_2} &
  (p',\sigma_2\sigma'_2) &  \xrightarrow{u_3/w_3} & (q',\sigma_2\sigma'_2) &
\xrightarrow{u_4/w_4} & (q',\sigma_2) \\
\end{array}\right.$
\item $|v_2v_4| \neq |w_2w_4|$, or $|v_2v_4| = |w_2w_4|$, 
  $|v_2v_4w_2w_4|\neq 0$, and there is a mismatch between $v_1v_3$ and 
  $w_1w_3$ (i.e. $v_1v_3 = x\alpha v$ and $w_1w_3 = x\beta w$ for some 
  $x,v,w\in\Sigma^*$ and some $\alpha\neq \beta\in\Sigma$). 
  \end{enumerate}
\end{lem}

\begin{proof}
Suppose that the MTP is not satisfied. By definition of the
MTP, there exist words and states that satisfy conditions~$(1)$ and~$(2)$, 
and such that $\Delta(v_1v_3,w_1w_3)\neq 
\Delta(v_1v_2v_3v_4,w_1w_2w_3w_4)$. If $|v_2v_4|\neq |w_2w_4|$ then we 
are done. Let us assume that $|v_2v_4| = |w_2w_4|$. Clearly, if $|v_2v_4w_2w_4| = 0$, then
 $\Delta(v_1v_3,w_1w_3) = \Delta(v_1v_2v_3v_4,w_1w_2w_3w_4)$ which is
 impossible. By Lemma~\ref{lem:diffcont}, the set
$\{ \Delta(v_1{(v_2)}^iv_3{(v_4)}^i,w_1{(w_2)}^iw_3{(w_4)}^i)\mid i\geq 0\}$ 
is infinite. We show that there exists $i_0$ such that there is a
mistmatch between $v_1(v_2)^{i_0}v_3(v_4)^{i_0}$ and
$w_1(w_2)^{i_0}w_3(w_4)^{i_0}$. Suppose that no such $i_0$ exists and that
$|v_1v_3|\leq |w_1w_3|$ (the other case is symmetric). 
Then, since $|v_2v_4| = |w_2w_4|$, for all $i\geq 0$, we have 
$|v_1{(v_2)}^iv_3{(v_4)}^i| =  |w_1{(w_2)}^iw_3{(w_4)}^i| - |w_1w_3|+|v_1v_3|$, and  $v_1{(v_2)}^iv_3{(v_4)}^i \preceq 
w_1(w_2)^iw_3(w_4)^i$. In other words, for all $i$, there exists
$x_i\in\Sigma^*$ such that $w_1(w_2)^iw_3(w_4)^i =
v_1(v_2)^iv_3(v_4)^ix_i$. Rephrased with delays, it means that
$\Delta(v_1(v_2)^iv_3(v_4)^i,w_1(w_2)^iw_3(w_4)^i) = (\epsilon,x_i)$
for all $i\geq 0$. However, since $|v_1(v_2)^iv_3(v_4)^i| =  |w_1(w_2)^iw_3(w_4)^i| - 
|w_1w_3|+|v_1v_3|$ for all $i\geq 0$, we have $|x_i| = 
|w_1w_3|-|v_1v_3|$ for all $i$. It contradicts the fact that
$\{ \Delta(v_1(v_2)^iv_3(v_4)^i,w_1(w_2)^iw_3(w_4)^i)\mid i\geq 0\}$ 
is infinite. Therefore, there exists $i_0\geq 0$ such that there is a 
mismatch between $v_1(v_2)^{i_0}v_3(v_4)^{i_0}$ and
$w_1(w_2)^{i_0}w_3(w_4)^{i_0}$. Finally, using the following
decomposition, one satisfies the three conditions of the Lemma
(i.e. by replacing $u_3$ by $u_2^{i_0}u_3(u_4)^{i_0}$ and so on):
\[
\begin{array}{l@{\hspace{0.3mm}}l@{\hspace{0.3mm}}l@{\hspace{0.3mm}}l@{\hspace{0.3mm}}l@{\hspace{0.3mm}}l@{\hspace{0.3mm}}l@{\hspace{0.3mm}}l@{\hspace{0.3mm}}llllllllll}
  (i,\bot) & \xrightarrow{u_1/v_1} & (p,\sigma_1) &  \xrightarrow{u_2/v_2} &
  (p,\sigma_1\sigma_1') &  \xrightarrow{{(u_2)}^{i_0}u_3{(u_4)}^{i_0}/{(v_2)}^{i_0}v_3{(v_4)}^{i_0}} & (q,\sigma_1\sigma'_1) &
\xrightarrow{u_4/v_4} & (q,\sigma_1) \\

  (i',\bot) & \xrightarrow{u_1/w_1} & (p',\sigma_2) &  \xrightarrow{u_2/w_2} &
  (p',\sigma_2\sigma_2') &  \xrightarrow{{(u_2)}^{i_0}u_3{(u_4)}^{i_0}/{(w_2)}^{i_0}w_3{(w_4)}^{i_0}} & (q',\sigma_2\sigma'_2) &
\xrightarrow{u_4/v_4} & (q',\sigma_2) \\
\end{array}
\]


\textbf{Conversely}, suppose that the conditions $(1)$, $(2)$ and
$(3)$ are satisfied. If $|v_2v_4|\neq |w_2w_4|$, then clearly,
the set $\{ \Delta(v_1{(v_2)}^{i}v_3{(v_4)}^i,w_1{(w_2)}^{i}w_3{(w_4)}^i)\mid
i\geq 0\}$ is infinite, and therefore there exist $i_0,i_1$ such that $i_0<i_1$ and
$\Delta(v_1{(v_2)}^{i_0}v_3{(v_4)}^{i_0},w_1{(w_2)}^{i_0}w_3{(w_4)}^{i_0})$ is
different from
$\Delta(v_1{(v_2)}^{i_1}v_3{(v_4)}^{i_1},w_1{(w_2)}^{i_1}w_3{(w_4)}^{i_1})$. Then,
the following decomposition witnesses the non-satisfiability of the MTP\@:

\noindent
\scalebox{0.85}{
$
\begin{array}{l@{\hspace{0.3mm}}l@{\hspace{0.3mm}}l@{\hspace{0.3mm}}l@{\hspace{0.3mm}}l@{\hspace{0.3mm}}l@{\hspace{0.3mm}}l@{\hspace{0.3mm}}l@{\hspace{0.3mm}}llllllllll}
  (i,\bot) & \xrightarrow{u_1/v_1} & (p,\sigma_1) &  \xrightarrow{u_2^{i_1{-}i_0}/v_2^{i_1{-}i_0}} &
  (p,\sigma_1{(\sigma_1')}^{i_1{-}i_0}) &  \xrightarrow{u_2^{i_0}u_3u_4^{i_0}/v_2^{i_0}v_3v_4^{i_0}} & (q,\sigma_1{(\sigma'_1)}^{i_1{-}i_0}) &
\xrightarrow{u_4^{i_1-i_0}/v_4^{i_1{-}i_0}} & (q,\sigma_1) \\

  (i',\bot) & \xrightarrow{u_1/w_1} & (p',\sigma_2) &  \xrightarrow{u_2^{i_1{-}i_0}/w_2^{i_1{-}i_0}} &
  (p',\sigma_2{(\sigma_2')}^{i_1{-}i_0}) &  \xrightarrow{u_2^{i_0}u_3u_4^{i_0}/w_2^{i_0}w_3w_4^{i_0}} & (q',\sigma_2{(\sigma'_2)}^{i_1{-}i_0}) &
\xrightarrow{u_4^{i_1-i_0}/w_4^{i_1{-}i_0}} & (q',\sigma_2) \\
\end{array}
$
}

\medskip
\noindent
Now, suppose that $|v_2v_4| = |w_2w_4|$, $|v_2v_4w_2w_4|\neq 0$ and
there is a mistmatch between $v_1v_3$ and $w_1w_3$, i.e.
$v_1v_3 = x\alpha v$ and $w_1w_3 = x\beta w$ for some
$x,v,w\in\Sigma^*$ and some $\alpha\neq \beta\in\Sigma$. Then
$\Delta(v_1v_3,w_1w_3) = (\alpha v, \beta w)$. Suppose that
$\Delta(v_1{(v_2)}^{i}v_3{(v_4)}^i,w_1{(w_2)}^{i}w_3{(w_4)}^i) = (\alpha v, \beta
w)$ for all $i\geq 0$.
\end{proof}

\begin{lem}\label{lem:decidability}
  The matched twinning property is decidable in \conptime for \fvpts.
\end{lem}

\begin{proof}
  The proof is very similar to the proof of Proposition~\ref{prop:HTP} for HTP\@.
  From an \fvpt $T$, we also construct in polynomial time a pushdown
  automaton with a constant number of (one reversal) counters
  that accepts any word $u=u_1u_2u_3u_4u_5$ such that
  $u_1u_2u_3u_4$ satisfies the premise of the \MTP but such that
  $\del(v_1v_3,w_1w_3) \neq \del(v_1v_2v_3v_4,w_1w_2w_3w_4)$ (\ie, the
  \MTP is not verified).  Therefore the \MTP holds if and only if no word
  is accepted by the automaton, and this can be checked in \conptime~\cite{vpts10}.

  As for HTP, the automaton simulates any two runs and guesses the decomposition
  $u_1u_2u_3u_4$. It checks that each run is in the same state after
  reading $u_1$ and $u_2$, and in the same state after reading $u_3$ and $u_4$.
  Using its stack, it verifies that $u_3$ and $u_2u_4$ are well-nested.
  With two additional counters it checks that $|v_2v_4|=|w_2w_4|$,
  if it is not the case then it accepts $u$ (the \MTP is not verified).
  Finally the automaton checks that $\del(v_1v_3,w_1w_3) \neq
  \del(v_1v_2v_3v_4,w_1w_2w_3w_4)$.
  This is done using the characterization given in
  Lemma~\ref{lem:mtpcharac} (item 3 in particular),
  with the technique described in the proof of Proposition~\ref{prop:HTP}.
\end{proof}

\subsection{Deciding OBM}


We show in this section that the \MTP and OBM coincide:
any twinned \fvpt can be evaluated with online bounded memory,
and OBM transductions can only be realized by twinned \fvpts.
In particular, this shows that being twinned is not only a property of
the transducer, but also of the transduction it defines.
Another consequence is that OBM is decidable for \fvpts.

\begin{thm}\label{thm:obm}
  Let $T$ be an \fvpt.
  $\inter{T}$ is OBM iff the $T$ is twinned, which is decidable
  in \conptime.
  In this case, the Turing transducer $\evalTT(T)$ runs,
  on an input stream $u$, in space
  complexity quadratic in the height of $u$.
\end{thm}

\begin{proof}
  Using Proposition~\ref{prop:MTP-implies-OBM} and
  Proposition~\ref{prop:OBM-implies-MTP} (both proved in this section),
  an \fvpt $T$ is twinned iff $\inter{T}$ is OBM\@.
  The former is decidable in \conptime by Lemma~\ref{lem:decidability}.
  Proposition~\ref{prop:complexity-eval} proves that the
  algorithm \eval uses at most
  $O((\hc(u')+1)\cdot\maxoutdiff(u'))$ space on the working tape
  after reading a prefix $u'$ of $u$.
  Proposition~\ref{prop:MTP-implies-OBM} shows that
  \[\maxoutdiff(u')\leq
  (\hc(u')+1) \cdot \left( 3(|Q|^{2(|Q|^4+1)}-1) \right)\cdot M\]
  when $T$ is twinned.
  Thus the space used by the algorithm is quadratic in $\hc(u')$.
\end{proof}

Twinned \fvpts define a class of transductions (and not just a class
of transducers):

\begin{cor}
Let $T,T'$ be two equivalent \fvpts. Then $T$ is twinned iff $T'$ also is.
\end{cor}

\begin{proof}
  By Theorem~\ref{thm:obm},
  $T$ is twinned iff $\inter{T}$ is OBM iff $\inter{T'}$ is OBM
  iff $T'$ is twinned.
\end{proof}

\begin{prop}\label{prop:MTP-implies-OBM}
  If an \fvpt $T$ is twinned then $\inter{T}$ is in OBM\@.
\end{prop}

\begin{figure}[t]
\centering

\begin{tikzpicture}[scale=.1]

  \draw (0,0)
  to [out=35,in=215] (7,11) node(pq1){$\bullet$} node[below right]{$(p,q)$}
  to [out=35,in=180] (15,25)
  to [out=0,in=180] (20,18)
  to [out=0,in=250] (30,32) node(pq2){$\bullet$} node[above left]{$(p,q)$}
  to [out=70,in=180] (35,38)
  to [out=0,in=120] (40,32) node(ppqp1){$\bullet$} node[above right]{$(p',q')$}
  to [out=290,in=120] (45,11) node(ppqp2){$\bullet$} node[above right]{$(p',q')$}
  to [out=310,in=180] (50,8)
  to [out=0,in=180] (65,20)
  to [out=0,in=120] (80,5)
  to [out=300,in=120] (90,0);

  \draw[->] (0,0) -- (0,40);
  \draw[->] (0,0) -- (95,0);
  \draw (0,42) node{height};
  \draw (100,0) node{input};

  \draw[dashed] (0,32) node[left]{$j$}  -- (40,32);
  \draw[dashed] (0,11) node[left]{$j'$} -- (45,11);
  \draw[dotted] (7, 0) node[below]{$\Left{j'}$}  -- (7,11);
  \draw[dotted] (45,0) node[below right]{$\Right{j'}$} -- (45,11);
  \draw[dotted] (30,0) node[below]{$\Left{j}$}  -- (30,32);
  \draw[dotted] (40,-6) node[below]{$\Right{j}$} -- (40,32);

\end{tikzpicture}

\caption{Vertical pumping in a well-nested word}\label{fig:vertical}
\end{figure}
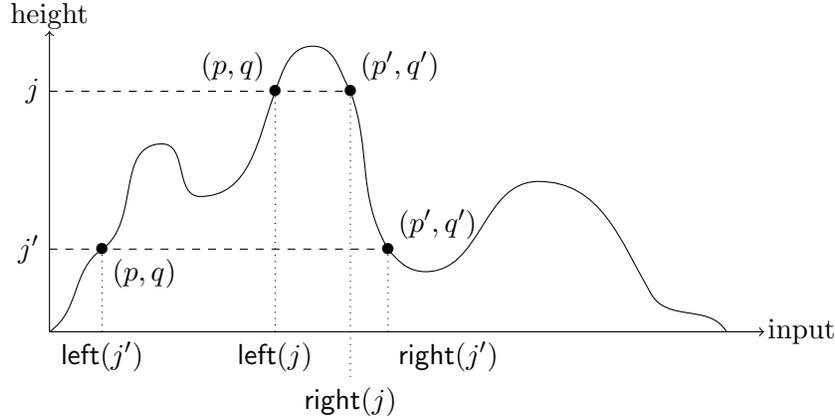

\begin{proof}
  \ignore{
  Sketch:
Like for the HTP, when $T$ satisfies the MTP, also does the reduced VPT returned by the
reduction procedure of~\cite{VPAreduction}.
We use a pumping technique to show
  that for any word $u\in\Sigma^*$ on which there is a run of $T$, we
  have $\maxoutdiffT(u)\leq (h(u)+1)q(T)$ for some function $q$,
  whenever the \MTP holds for $T$.  This is done as follows: any such
  word can be uniquely decomposed as $u = u_0c_1u_1c_2\dots c_{n}u_{n}$
  with $n\leq h(u)$, each $u_i$ is well-nested and each $c_i$ is a
  call. Then if the $u_i$ are long enough, we can pump them vertically
  and horizontally without affecting the global delay, by using the
  property $\Delta(vv',ww') = \Delta(\Delta(v,w).(v',w'))$. Then we
  can apply Proposition~\ref{prop:complexity-eval}.

} 

\olivier{A relire}
  Let $T$ be a twinned \fvpt.
  We show that, for all input words $u$ and all prefixes $u'$ of $u$,
  \[\maxoutdiff(u')\leq
  (\hc(u')+1) \cdot \left( 3(|Q|^{2(|Q|^4+1)}-1) \right)\cdot M,\]
  where $M$ is the length of the longest output occurring on the
  transitions of $T$.   Using Proposition~\ref{prop:complexity-eval},
  this shows that $\inter{T}$ is in OBM\@.

  Let $u\in\dom(T)$, and $u'$ be a prefix of $u$.
  %
  There exists a unique decomposition of $u'$ as follows:
  $
  u' = u_0 c_1 u_1 c_2 \ldots u_{n-1} c_n u_n
  $,
  where $n=\hc(u')$ (the current height of $u'$),
  and for any $i$, $c_i\in \Sigma_c$ and
  $u_i$ is well-nested.
  %
  If each of the $u_i$'s is such that $|u_i| \leq
  3(|Q|^{2(|Q|^4+1)}-1)$, then the property holds as the length of
  $u'$ can be bounded by \[(\hc(u')+1) \cdot  \left (
    3(|Q|^{2(|Q|^4+1)}-1) \right ).\]

  Otherwise, we prove that there exists a strictly shorter input word
  that produces the same delays as $u'$ when evaluating the
  transduction on it.
  If $|Q|=1$ then we can apply the HTP (implied by the MTP)
  on any of the $u_i$ to show that
  removing it would not change the delays
  (as in the proof of Lemma~\ref{lm:HTP-delay}).
  If every $u_i$ is empty, then $|u'|=\hc(u')$ and we get the result.
  Now, assume that $|Q|>1$.
  Let $(q,\sigma,w)$, $(q',\sigma',w')\in Q\times \Gamma^*\times \Sigma^*$
  be such that there exist runs
  $\rho:  (q_0,\bot)\xrightarrow{u'/v}(q,\sigma)$ and
  $\rho': (q_0',\bot)\xrightarrow{u'/v'}(q',\sigma')$, with $q_0,q_0'\in I$,
  $v = \lcpin {u'} T\cdot w$, $v' = \lcpin {u'} T\cdot w'$,
  and such that $\maxoutdiff(u') = |w|$.
  Consider the smallest index $i$ such that $|u_i'| >
  3(|Q|^{2(|Q|^4+1)}-1)$. We distinguish two cases:
  \begin{enumerate}
  \item if $\h(u_i)\leq |Q|^4$,
    then $u_i'$ is not $|Q|^2$-narrow.
    Indeed, if it was, we could apply Lemma~\ref{lem:arity-bound} (as $|Q|>1$)
    and get $|u_i|\le 3(|Q|^{2(\h(u_i)+1)}-1) \le 3(|Q|^{2(|Q|^4+1)}-1)$,
    a contradiction.
    So $u_i$ is not $|Q|^2$-narrow.
    In this case we ``pump horizontally''
    like in the proof of Lemma~\ref{lm:HTP-delay}
    as there exists a well-nested factor $u''$ from which we can remove
    a non-empty factor while preserving the delays.
  \item if $\h(u_i) > |Q|^4$, we prove that we can ``pump
    vertically'' $u_i$, and thus reduce its length too. Indeed, let
    $k$ be the first position in word $u_i$ at which height $\h(u_i)$
    is obtained. As $u_i$ is well-nested, we can define for each
    $0\leq j < \h(u_i)$ the unique position $\Left{j}$
    (resp. $\Right{j}$) of $u_i$ as the largest index, less than $k$ (resp.\ the
    smallest index, larger than $k$), whose height is $j$ (see
    Figure~\ref{fig:vertical}). As $\h(u_i) > |Q|^4$, there exist two
    heights $j$ and $j'$ such that configurations reached at positions
    $\Left{j}$, $\Left{j'}$, $\Right{j}$ and $\Right{j'}$ in runs
    $\rho$ and $\rho'$ satisfy the premises of the matched twinning
    property, considering a prefix of $u_0 c_1 \ldots c_i u_i$.
    Thus, one can replace in this prefix $u_i$ by a shorter word
    $u'_i$ and hence reduce its length, while preserving the delays
    reached after it. Let $u''$ be the word obtained from $u'$ by
    substituting $u'_i$ to $u_i$, hence $|u''|<|u'|$. By
    Lemma~\ref{lem:assocdelay}, this entails that the delays reached
    after $u'$ and $u''$ are the same, proving the result.
    \qedhere
  \end{enumerate}
\end{proof}

\begin{prop}\label{prop:OBM-implies-MTP}
  Let $T$ be an \fvpt. If $\inter{T}$ is in OBM then $T$ is twinned.
\end{prop}

\begin{proof}
  Consider a \vpt $T$ that does not satisfy the \MTP,
  and assume for contradiction that there exists
  an OBM Turing transducer $A$
  computing the transduction of $T$.
  As $T$ does not satisfy the MTP, we can find two runs as in the definition
  of the MTP, that accumulate different delays:
\[
\begin{array}{llllllllllllllllll}
  (i,\bot) & \xrightarrow{u_1/v_1} & (p,\sigma_1) &  \xrightarrow{u_2/v_2} &
  (p,\sigma_1\sigma_1') &  \xrightarrow{u_3/v_3} & (q,\sigma_1\sigma'_1) &
\xrightarrow{u_4/v_4} & (q,\sigma_1) \\

  (i',\bot) & \xrightarrow{u_1/w_1} & (p',\sigma_2) &  \xrightarrow{u_2/w_2} &
  (p',\sigma_2\sigma'_2) &  \xrightarrow{u_3/w_3} & (q',\sigma_2\sigma'_2) &
\xrightarrow{u_4/w_4} & (q',\sigma_2) \\
\end{array}
\]

\smallskip
\noindent and $\del(v_1v_3,w_1w_3) \neq
\del(v_1v_2v_3v_4,w_1w_2w_3w_4)$.

Remind that in the definition of MTP, we require that $(q,\sigma_1)$ and
$(q',\sigma_2)$ are co-accessible configurations. Therefore there exists
two continuations $u_5$ and $u'_5$ that lead to accepting configurations from
these two configurations respectively. Assume that the runs on $u_5$ and $u_5'$
produce two output words $v_5$ and $w_5$ respectively.

For all $\alpha\in\mathbb{N}$, we let
\[
\begin{array}{llllllll}
    U_\alpha & = & u_1u_2^\alpha u_3u_4^\alpha \\
    V_\alpha & = & v_1v_2^\alpha v_3v_4^\alpha \\
    W_\alpha & = & w_1w_2^\alpha w_3w_4^\alpha \\
\end{array}
\]

Note that $U_\alpha u_5$ and $U_\alpha u'_5$ are both accepted for all $\alpha$, and
their respective outputs are $V_\alpha v_5$ and $W_\alpha w_5$.

\manu{I have rewritten this below, it was not precise}
Now, after reading $U_{\alpha}$ the stack height $h\in\mathbb{N}$ is always the same
for all $\alpha$. Since $A$ is OBM, the amount of information written
on its working tape after reading $U_\alpha$ only depends on $h$.
Hence, there exists a pair $(c,q)$ where $c$ is a
word written on the working tape of $A$ and $q$ is a state of $A$,
such that for infinitely many $\alpha$, $A$ is in the configuration
$(c,q)$ after reading $U_\alpha$. Let $\Xi$ denote the set of such $\alpha$'s.

We denote by $V$ the output produced by $A$ on $u_5$ from the configuration $(c,q)$,
and by $W$ the output produced by $A$ on $u'_5$ from $(c,q)$.
For all $\alpha$, we also denote by $O_\alpha$ the output of $A$ on $U_\alpha$
(it is unique since the machine is deterministic).
Therefore, for all $\alpha\in \Xi$, we have:
\[
\begin{array}{lllllll}
\inter{T}(U_\alpha u_5) & = & O_\alpha V = V_\alpha v_5\\
\inter{T}(U_\alpha u'_5) & = & O_\alpha W = W_\alpha w_5 \\
\end{array}
\]

Hence $O_\alpha = V_\alpha v_5 V^{-1} = W_\alpha w_5 W^{-1}$
for all $\alpha \in \Xi$. We show that it implies that there exists a
delay $d$ and infinitely many $\alpha$ such that
$\del(V_\alpha,W_\alpha) = d$, which contradicts
Lemma~\ref{lem:diffcont} below (applied to $m=n=2$) since
$\del(V_0,W_0)\neq \del(V_1,W_1)$ by our initial assumption.

Suppose that it is not the case, then we can find an arbitrarily large
delay between $V_\alpha$ and $W_\alpha$. We then choose $\alpha$ such
that $\del(V_\alpha,W_\alpha) = (x,y)$ with
$\text{max}(|x|,|y|)>|V|+|W|+|v_5|+|w_5|$. It
means that $V_\alpha = \ell x$ and $W_\alpha = \ell y$ for some
$\ell$. Then, we know that $V_\alpha v_5 V^{-1} = W_\alpha w_5
W^{-1}$, hence $x v_5 V^{-1} = y w_5 W^{-1}$. We also have
$|x|+|v_5|-|V| \leq |xv_5V^{-1}|\leq |x|+|v_5|+|V|$, and
$|y|+|w_5|-|W| \leq |yw_5W^{-1}|\leq |y|+|w_5|+|W|$. We distinguish
three cases:
\begin{itemize}
    \item $x = \epsilon$: then $|y|+|w_5|-|W|\leq |yw_5W^{-1}| =
      |xv_5V^{-1}|\leq |v_5|+|V|$, and $|y|\leq |v_5|+|V|+|W|-|w_5|$
      which contradicts $|y|> |v_5|+|V|+|W|+|w_5|$,
    \item $y = \epsilon$: this case is symmetric to the previous one,
    \item $x = ax'$ and $y=by'$ for some letters $a\neq b$: then we
      have $ax' v_5 V^{-1} = by' w_5 W^{-1}$, which implies that
      either the first $a$ is ``erased'' by $V^{-1}$ or the first $b$
      is erased by $W^{-1}$, but it cannot be the case that both these
      letters are erased due to the length of $x$ and
      $y$. Suppose that $a$ is erased by $V^{-1}$, then $ax'v_5V^{-1}$
      is of the form $\beta^{-1}$ for some $\beta$ (it is an inverse word), while
      $by'w_5W^{-1}$ is not. The other case is symmetric.
      \qedhere
\end{itemize}
\end{proof}

\begin{rem}[Sequential \vpts]
Sequential transducers have at most one run per input word, so
\emph{sequentializable \vpts are twinned}.
The \MTP is not a sufficient condition to be sequentializable, as
shown for instance by Example~\ref{exple:vpt3}.
Therefore the class of transductions defined by transducers which
satisfy the \MTP is strictly larger than the class of transductions
defined by sequentializable transducers. However,
these transductions are
in the same complexity class for evaluation, i.e., polynomial space in
the height of the input word for a fixed transducer.
\end{rem}

\ignore{
\begin{thm}\label{thm:polyhbm}
  Let $T$ be an \fvpt. If $T$ is
  twinned, then the Turing transducer $\evalTT(T)$ runs, on an input
  stream $u$, in space complexity quadratic in the height of $u$.
\end{thm}


} 


\section{Conclusion and Remarks}%
\label{sec:conclusion}

\manu{j'ai mis a jour}
This work investigates the streaming evaluation of nested word
transductions defined by visibly pushdown transducers. The main
result is the introduction of two classes of \vpt-transductions, shown
to be decidable (in the class of \vpt-transductions):
\vpt-transductions which can be evaluated in streaming with a memory that depends
only on the height of the input nested word (HBM) and on the current height
of the prefixes of the input nested word (OBM), respectively. These
two classes have been effectively characterized by structural
properties of \vpts, respectively called horizontal and matched
twinning properties. We have
designed a streaming algorithm to evaluate \vpt in general and
analysed its space complexity. This algorithm, applied to a \vpt
satisfying the horizontal twinning property, runs in height bounded
memory. Applied to a \vpt satisfying the matched twinning property, it
runs in online bounded memory.

The following inclusions summarize the relations between the different \emph{classes} of
transductions we have studied:
\[\text{BM \fvpts}\subsetneq \text{Sequentializable \vpts}\!\subsetneq\! \text{OBM \fvpts}\subsetneq
\text{HBM \fvpts}\! \subsetneq\! \text{\fvpts}
\]
 Moreover, we have shown that BM, OBM and HBM \fvpts are decidable
 in \conptime.



\subsection*{Further Directions}
 An important asset of the class of
OBM \fvpts w.r.t.\ the class of sequentializable \vpts is that
it is decidable. It would thus be interesting to determine
whether or not the class of  sequentializable \vpts is decidable, and
to characterize the class of sequentializable \vpts in terms of memory
requirements.
%
%
In addition, we also plan to extend our techniques to more expressive
transducers, such as two-way visibly pushdown transducers as
introduced in~\cite{DBLP:conf/lics/DartoisFRT16}, which are equivalent
to MSO-transducers from nested words to words.  For (flat) words,
deciding bounded memory of a transduction given by a finite transducer
amounts fo decide whether it is sequentializable. If the transduction
is given by a two-way (flat) word transducer, or equivalently by an
MSO-transducer~\cite{EngelfrietH01}, deciding bounded memory can be
done by first checking whether the transduction is rational,
i.e., whether it is realizable by a (one-way) finite state transducer,
and then by deciding sequentializability of the one-way
transducer. The first step has been shown to be decidable in~\cite{DBLP:conf/lics/FiliotGRS13},
with an elementary complexity in~\cite{bgmp17}.

To extend the result of this paper to other models of transducers, say
two-way visibly pushdown transducers, we plan to extend the result of~\cite{DBLP:conf/lics/FiliotGRS13} to nested words. I.e., given a
two-way visibly pushdown transducer, decide whether it is equivalent
to some \vpt.

Another line of work concerns the extension of our evaluation
procedure beyond functional transductions, or to multi-input and
multi-output transductions.

\section*{Acknowledgement}
The authors would like to
thank Jean-Fran\c{c}ois Raskin and Stijn Vansummeren for their
comments on a preliminary version of this work.


\bibliographystyle{alpha}
\bibliography{main}




\end{document}